\documentclass[12pt]{article}
\RequirePackage{amsthm,amsmath,amsfonts,amssymb}
\RequirePackage[authoryear]{natbib}
\RequirePackage[colorlinks,citecolor=blue,urlcolor=blue]{hyperref}
\usepackage[utf8]{inputenc}
\usepackage{mathrsfs}
\usepackage{graphicx,psfrag,epsf, graphics}
\usepackage{psfrag,pst-node,subfigure,rotating,  picture, epsfig, upgreek}
\usepackage{enumerate}
\usepackage{cases}
\usepackage{url} 
\usepackage{epstopdf}
\usepackage{extarrows}
\usepackage{threeparttable}
\usepackage{tabularx}
\usepackage{makecell}
\usepackage{multirow}
\usepackage{caption}
\usepackage{booktabs} 
\usepackage{ulem}
\usepackage{xr}
\numberwithin{equation}{section}
\theoremstyle{plain}

\newtheorem{theorem}{Theorem}[section]
\newtheorem{lemma}{Lemma}[section]

\newtheorem{proposition}{Proposition}[section]

\newtheorem{corollary}{Corollary}[section]

\theoremstyle{remark}
\newtheorem{remark}{Remark}

\newtheorem{assumption}{Assumption}
\def\y{\mathbf y}
\def\0{\mathbf 0}

\def\d{\mathbf d}

\def\h{\mathbf h}
\def\f{\mathbf f}

\def\w{\mathbf w}
\def\A{\mathbf A}

\def\Q{\mathbf Q}

\def\W{\mathbf W}
\def\I{\mathbf I}
\def\S{\mathbf S}

\def\K{\mathbf K}
\def\H{\mathbf H}
\def\R{\mathbf R}
\def\E{\mathbb E}
\def\M{\mathbf M}
\def\N{\mathrm N}
\def\W{\mathbf W}

\def\x{\mathbf x}

\def\f{\mathbf f}
\def\T{\mathrm T}

\def\be{\begin{equation}}
\def\ee{\end{equation}}
\def\bea{\begin{eqnarray}}
\def\eea{\end{eqnarray}}
\def\nn{\nonumber}
\def\bfeta{\boldsymbol {\eta}}

\def\bfep{\boldsymbol \varepsilon}

\def\bfSigma{\boldsymbol \Sigma}

\def\bfXi{\boldsymbol \Xi}

\def\fn{\footnote}

\def\addwsx{\textcolor{blue}}
\def\hd{high dimensional}

\newcommand{\blind}{1}

\begin{document}
\date{}

\def\spacingset#1{\renewcommand{\baselinestretch}%
{#1}\small\normalsize} \spacingset{1}


\if1\blind
{
  \title{\bf High Dimensional Ensemble Kalman Filter}
  \author{
    Shouxia Wang\\
 \small  School of Statistics and Management, Shanghai University of Finance and Economics
  \\
  Hao-Xuan Sun
  \\
  \small Center for Big Data Research, Peking University\\
  and\\
Song Xi Chen\thanks{}\\ 
\small Department of Statistics and Data Science, Tsinghua University }
  \maketitle
} \fi

\if0\blind
{
  \bigskip
  \bigskip
  \bigskip
  \begin{center}
    {\LARGE\bf High Dimensional Ensemble Kalman Filter}
\end{center}
  \medskip
} \fi

\bigskip
\begin{abstract}
The Ensemble Kalman Filter (EnKF), as a fundamental data assimilation approach, has been widely used in many fields of the sciences and engineering. When the state variable is of high dimensional accompanied with high resolution observations of physical models, some key theoretical aspects of the EnKF are open for investigation. 
This paper proposes several high dimensional EnKF (HD-EnKF) methods equipped with consistent estimators for the important forecast error covariance and Kalman Gain matrices.  It then studies the theoretical properties of the EnKF under both  fixed and high dimensional state variables, which provides one-step and multiple-step mean square errors of the analysis states to the underlying oracle states offered by the Kalman Filter and gives the much needed insight to the roles played by the forecast error covariance on the accuracy of the EnKF.  The accuracy of the data assimilation under the misspecified physical model is also considered.  Numerical studies on the Lorenz-96 and the Shallow Water Equation models illustrate that the proposed HD-EnKF algorithms outperform  the standard EnKF and widely used inflation methods.
\end{abstract}

\noindent%
\vfill

\newpage
\spacingset{1.9} 
\section{Introduction}
\label{sec-introduction} 

  Kalman Filter \citep{kalmanNewApproachLinear1960}
  is a fundamental approach for forecasting and data assimilation for a  partially observed dynamic system.  
 Data assimilation is a statistical procedure that 
 combines information from numerical physical model forecasts with observed data to obtain the best possible description of a dynamical system and its uncertainty \citep{evensen2022}. The assimilated state variables at a time are used to generate the forecast for the next time step.  For chaotic systems sensitive to initial values, data assimilation can significantly improve the accuracy of the forecast by using optimized state variables as initial values \citep{Lorenz1963}.
  Although 
much applied in numerical weather and oceanography prediction,  Kalman Filter and its later developments (EnKF, three and four-dimensional variational assimilation (3D-Var, 4D-Var; \citealp{pailleux1990global,courtierECMWF1998})) are the methodological core for scientific algorithms over various fields, including hydrology, remote sensing data inversion, carbon data assimilation, robotic control and automatic driving.
 Data assimilation methods also provide the underlying algorithms and predictive techniques for building digital twins for social and economic systems. 
 
The limitations of the Kalman Filter are (i) 
 it assumes a linear model which is restrictive for many non-linear 
  dynamic systems; and (ii) 
 the forecast error covariance and Kalman Gain matrices are expensive to store and update when the dimension of the state vector is large. To overcome these limitations, \cite{evensenSequentialDataAssimilation1994} proposed the Ensemble Kalman Filter (EnKF),  where the forecast error covariance is estimated by the sample covariance based on generated forecast ensembles.  The EnKF only requires storing and operating on much reduced-size
 state vectors and the ensembles can be generated in parallel.
 These advantages have made the EnKF an effective and widely used data assimilation scheme. 

Estimating the forecast error covariance is a key in 
the EnKF, 
which is made in the EnKF 
 by the sample covariance of the forecast ensembles. 
 Past works on the  EnKF found that the sampling error can lead to underestimation of the forecast error covariance and can eventually cause filter divergence\citep{andersonMonteCarloImplementation1999,constantinescuEnsemblebasedChemicalData2007}.
 Filter divergence refers to a decreasing ability 
  to correct the ensemble state towards the observations 
  after certain rounds of assimilation. 
One reason for the filter divergence is the variance of the ensemble forecast error becomes too small when there are unaccounted model errors \citep{Lorenc2003}; the other is due to the substantial mismatch between the estimated and true error covariance matrices 
\citep{houtekamerDataAssimilationUsing1998,constantinescuEnsemblebasedChemicalData2007}. 

Along with advances in remote sensing technology in measuring the Earth system, 
 there is a growing availability of high spatio-temporal resolution satellite observations \citep{stroudEnsembleKalmanFilter2010,guinehutHighResolution3D2012,RN768}. Meanwhile, the high resolution dataset  and forecasting are in increasing demands in modern meteorology and atmospheric systems like high resolution weather forecasting, global carbon calculation and ocean data assimilation, which leads to high dimensional geophysical models with high dimensional state vectors.  
The high dimensional state vectors lead to  \hd\  forecast error covariance and Kalman Gain matrices. The high dimensionality is known to cause the non-convergence of the sample forecast error covariance \citep{bai1993}, felt in the form of filter divergence and unreliable data assimilation in practice.   Geophysicists have proposed two methods to counter the problems,  the localization method 
\citep{furrerCovarianceTaperingInterpolation2006,furrerEstimationHighdimensionalPrior2007}  
   that sets covariances to zero for the geographical distances bigger than a threshold   
   and the inflation method that  
 enlarges the forecast error covariance by adding either a multiplicative factor or an additive term \citep{wangComparisonBreedingEnsemble2003a,constantinescuEnsemblebasedChemicalData2007,liangMaximumLikelihoodEstimation2012a}.
 However, the statistical properties of these methods are yet to be established. 
 \cite{tong2018performance} investigated the performance of the local EnKF with domain localization  under linear systems.  
 \cite{Katzfuss2020} proposed Gibbs EnKF and particle EnKF algorithms for high dimensional hierarchical dynamic models from a Bayesian perspective.
 %
  \cite{alghattas2024non} developed a non-asymptotic analysis on the  difference between the ensembles of single-step Ensemble Kalman Inversion using  the true and the localized estimation of forecast error covariances under the framework of effective  dimension of matrices and \cite{Ghattas2024siam} explored the resampling EnKF under linear systems and effective dimension. While \cite{alghattas2025covariance} studied 
 the difference between the ensembles of single-step EnKF using the true  and the thresholded 
 estimation of forecast error covariance functions  in a small correlation lengthscale regime. 
{The EnKF methods and their theoretical analysis under high dimensional  framework for general non-linear models, especially the theoretical performance of the multiple-step EnKF under both the correct and misspecified models, need further investigation.}  

This paper proposes several \hd\ EnKF (HD-EnKF) methods and conducts a theoretical analysis of the one-step and multiple-step assimilated state vectors for general nonlinear models and imperfect models under both fixed and high dimensional cases.  
It is shown that
even at the \hd\ setting, the mean square errors (MSEs)  of the analysis states by the EnKF could be controlled by the error bounds of the high dimensional forecast error covariance estimators. Thus, 
consistent recovery of the underlying analysis states can be attained by the proposed HD-EnKF. 
Specifically, we establish the MSE bounds and weak convergence of the analysis states from the HD-EnKF to the underlying analysis under both fixed and high dimensional cases, as well as for both correct and imperfect physical models. 
The MSE bounds for the multiple-step forecast and assimilation are also provided. 
The investigation of the HD-EnKF under the 
misspecified model gives a more realistic description of the performance of the EnKF.
{Under the misspecified model, the MSE not only depends on the estimation error of the forecast error covariance matrices but also the discrepancy between the misspecified and the true model.}  
Simulation studies 
suggest that the proposed HD-EnKF could achieve more accurate and robust assimilation results than the widely used inflation-based EnKF method. This implies that the proposed HD-EnKF is applicable in a range of studies and prospects in earth science like global carbon calculation, weather forecasting and the production of high resolution data sets, considering  both the theoretical guarantee and better performances in practice.   

The paper is structured as follows.
Section \ref{sec-2enkf} reviews the Kalman Filter and the EnKF. 
Section \ref{sec-enkfconvergefix} presents several 
consistent estimators of the high dimensional forecast error covariance and the key Kalman Gain, and establishes the theoretical properties of the EnKF. The effects of imperfect or misspecified models on the accuracy of the EnKF are provided in Section \ref{sec-5-imperfect-model}. Results of empirical data assimilation based on two well-known geophysical models are reported in Section \ref{simulation}, 
followed by a conclusion in Section \ref{sec-discussion}. 
The proposed HD-EnKF algorithms, additional theoretical and simulation results, and technical details
 are given in the Supplementary Material (SM).

    \section{Review on Kalman and Ensemble Kalman Filters} \label{sec-2enkf}
	The Kalman Filter \citep{kalmanNewApproachLinear1960}  
 assumes the following linear Gaussian state-space model (which was generalized to nonlinear models in the EnKF) 
 \begin{equation}\label{state-space}	    
		\left\{\begin{array}{l} 
			\mathbf{x}_t=\mathbf{M}_{t} \mathbf{x}_{t-1}+\mathbf{w}_t, \quad\ \quad\mathbf{w}_t \sim \mathrm{N}_p\left(\0, \mathbf{Q}_t\right), \\
			\mathbf{y}_{t}=\mathbf{H}_{t} \mathbf{x}_{t}+\bfep_{t}, \quad\quad\quad\quad\,\bfep_{t} \sim \mathrm{N}_q(\0, \mathbf{R}_t),
		\end{array}\right.
	\end{equation}
	where $\mathbf{x}_{t}$ is the  $p-$dimensional not fully observed state vector of a dynamic system at time $t$, $\mathbf{M}_{t}$ is the linear model operator (also called evolution or forecast operator), $\mathbf{w}_t$ is the model error term 
 with $\mathbf{Q}_t$ as its covariance, $\mathbf{y}_{t}$ is the $q-$dimensional observation vector at time $t$,  $\mathbf{H}_{t}$ is a linear observation operator which relates the state $\mathbf{x}_{t}$ to the observation  $\mathbf{y}_{t}$, and $\bfep_{t}$ is the observation error with $\mathbf{R}_t$ being its covariance matrix.  
 
 Assuming $\mathbf{M}_{t}$, $\mathbf{Q}_t$, $\mathbf{H}_{t}$, and $\mathbf{R}_t$ are 
 known, 
 the Kalman Filter provides a two-step alternative updating framework based on the theory of multivariate normal distribution. In the forecast step, suppose that $\x_{t-1} \mid \y_{1: t-1} \sim {\N}_p(\x_{t-1}^{a}, \bfSigma_{t-1}^a)$, where $\x_{t-1}^{a}$ is the analysis state vector at the previous time $t-1$ and $\y_{1: t-1}$ denotes the observation  $\{\mathbf{y}_i\}_{i=1}^{t-1}$ up to time $t-1$, the conditional distribution of $\x_{t}$ given $\y_{1: t-1}$ is 
	\begin{equation}
		\x_t \mid \y_{1: t-1} \sim {\N}_p\big(\x_t^f, \bfSigma_t^f\big) \equiv {\N}_p\big({\M}_{t} \x_{t-1}^{a}, {\M}_{t} \bfSigma_{t-1}^a {\M}_{t}^{\T}+\Q_t\big),
		\label{foreupdate}
	\end{equation}
	where $\mathbf{x}_t^{f}=\mathbf{M}_{t} \mathbf{x}_{t-1}^{a}$ is the forecast state and 
	\begin{equation}
		\bfSigma_t^{f}=\mathbf{M}_{t} \bfSigma_{t-1}^a  \mathbf{M}_{t}^{\mathrm{T}}+\mathbf{Q}_t, 
		\label{forecov}
	\end{equation}
	 is the forecast error covariance, the variance of the  forecast error 
   $\x_t-\x_t^f$. 
 
In the update (assimilation) step, the goal is to update $\mathbf{x}_{t}^{f}$ according to the new observation $\mathbf{y}_{t}$ and to reduce the difference between the true state $\x_t$ and the forecast state $\mathbf{x}_{t}^{f}$. The joint distribution of $\left(\x_t,\y_t\right)$ given  $\y_{1: t-1}$ is Gaussian,
	which implies that $ \x_t \mid \y_{1: t} \sim {\N}_p\left({\x}_t^{a}, {\bfSigma}_t^a\right)$,
    where,   by the conditional Gaussian distribution formulae, 
   \begin{align}	    	
			\x_t^{a}&={\x}_t^{f}+\K_t\big(\y_t-\mathbf{H}_t {\x}_t^{f}\big), \,\label{analysisupdate1}
				\bfSigma_t^a=\left(\I_p-\K_t \mathbf{H}_t\right) \bfSigma_t^f 
   \text{  and }\\
		\mathbf{K}_t&={\bfSigma}_t^{f} \mathbf{H}_t^{\mathrm{T}}\big(\mathbf{H}_t {\bfSigma}_t^{f} \mathbf{H}_t^{\mathrm{T}}+\mathbf{R}_t\big)^{-1},
		 \label{kalmangain}
\end{align}
which are, respectively, the analysis (assimilation) state and the associated analysis covariance matrix, and the Kalman Gain matrix. Note  that 
the analysis state $\x_t^{a}$ is the minimum variance unbiased estimator for the mean of $\x_t$ 
under the  linear state-space model \eqref{state-space}.	 

 However,  models 
 in earth science or engineering are often nonlinear
 so that $\mathbf{x}_t=\mathcal{M}_{t} (\mathbf{x}_{t-1})+\mathbf{w}_t$ in \eqref{state-space},
 where $\mathcal{M}_{t}$ is a nonlinear model operator that can 
 invalidate the forecast distribution in \eqref{foreupdate}.  Besides, for large dynamic systems, 
 the dimension of the state vector $\x_t$ can be large so that the forecast error covariance $\bfSigma_t^{f}$ is expensive to compute and store. To tackle these problems, \cite{evensenSequentialDataAssimilation1994} proposed the EnKF, where the forecast error covariance is no longer updated as \eqref{forecov} but by the sample covariance of $n$ forecast ensembles. It hence only needs to store and operate on $n$ vectors of dimension $p$,  and is convenient for parallel computing, and thus improves the computing efficiency and reduces the storage cost.

 The EnKF assumes possibly nonlinear model system with a linear observation operator, with the first equation in \eqref{state-space} replaced by  
    \begin{equation}\label{forecast-observation}	 	\mathbf{x}_t=\mathcal{M}_{t}\left(\x_{t-1}\right)+\w_t,
\end{equation} 
where $\mathcal{M}_{t}$ is the possibly nonlinear model operator.

A key feature of the EnKF is to generate an ensemble of  perturbed forecast states  $\{\x_{t, j}^f\}_{j=1}^n$ of size $n$ 
from the  conditional distribution of $\x_{t}$ given  $\y_{1: t-1}$, which are 
 used to  estimate the forecast error covariance matrix $\bfSigma_t^{f}$. 
 Given the analysis ensemble at the previous time step $\{\x_{t-1, j}^a\}_{j=1}^n$, 
 the EnKF consists of the following three steps.  
  	\begin{itemize}
		\item[Step (i):] Run the model forward in time to get the perturbed 
   forecast state and their average:
\begin{equation}\label{perturbforecast}
 \x_{t, j}^f=\mathcal{M}_{t}\left(\x_{t-1, j}^a\right) +\mathbf{w}_{t,j},\, \hat\x_t^f=\frac{1}{n}\sum_{j=1}^n \x_{t, j}^f,\,\mathbf{w}_{t,j} \stackrel{IID}\sim \mathrm{N}_p\left(\0, \mathbf{Q}_t\right).   
\end{equation}
		\item[Step (ii):] Estimate the forecast error covariance matrix $\bfSigma_t^f$ by the sample covariance matrix  of the forecast ensemble, namely 
		\begin{equation}\label{samplecov}
			\hat\bfSigma_t^f = (n-1)^{-1} \sum_{j=1}^n\big(\x_{t, j}^f-\hat\x_t^f\big)\big(\x_{t, j}^f-\hat\x_t^f\big)^{\T}\triangleq \S_t^f, 
		\end{equation} 
and obtain the Kalman Gain estimate $\hat\K_t=\hat\bfSigma_t^f \mathbf{H}_t^{\mathrm{T}}\big(\mathbf{H}_t \hat\bfSigma_t^f \mathbf{H}_t^{\mathrm{T}}+\mathbf{R}_t\big)^{-1}$. 
		\item[Step (iii):] Calculate the perturbed 
   residuals
		$
		\d_{t, j}=\y_t+\bfep_{t, j}-{\H}_t \x_{t, j}^f $ 
		where $\{\bfep_{t, j}\}_{j=1}^{n} \stackrel{IID}  \sim {\N}\left(\mathbf{0}, \R_t\right)$, and   
  let $\d_t=n^{-1} \sum_{j=1}^n \d_{t, j}$. Compute the perturbed
analysis states and their average, the analysis state, 
\begin{equation}\label{enkfupdate}
		\begin{aligned}
			\x_{t, j}^a&=\x_{t, j}^f+\hat\bfSigma_t^f \H_t^{\T}\big(\H_t \hat\bfSigma_t^f \H_t^{\T}+\R_t\big)^{-1} \d_{t, j},\\
			\hat\x_t^a&=\frac{1}{n} \sum_{j=1}^n \x_{t, j}^a=\hat\x_t^f+\hat\bfSigma_t^f \H_t^{\T}\big(\H_t\hat\bfSigma_t^f \H_t^{\T}+\R_t\big)^{-1} \d_t,
		\end{aligned}
    \end{equation}
which will be used for the data assimilation at the next time step. 
	\end{itemize} 

 In practice, the algorithm is initialized at time $t = 0$ by drawing $\{\x_{0, j}^a\}_{j=1}^n$ independently from ${\N}_p(\x_{0}^{a}, \Sigma_{0}^a)$ where $\x_{0}^{a}$ and $\Sigma_{0}^a$ are either chosen based on the physical model or by pre-running the physical model for a period of time.  
 Then,  the ensemble is propagated forward through time, alternating between the forecast and the analysis steps, according to the three-step procedure above. 
If $\R_t$ and $\Q_t$ are unknown, their estimates 
can be entertained to substitute them 
in \eqref{perturbforecast} and  \eqref{enkfupdate}.  

  The EnKF does not require the 
  expensive
 recursions \eqref{forecov} and \eqref{analysisupdate1}-\eqref{kalmangain} used in the Kalman Filter. Instead, it estimates the forecast error covariance $\bfSigma_t^{f}$ by the sample covariance. 
For the case with non-linear model operator as in \eqref{forecast-observation}, we can still define the underlying analysis state $\x_{t}^{a}$  by \eqref{analysisupdate1}-\eqref{kalmangain}, where $\x_t^f=\E\left(\x_{t}\mid \y_{1: t-1}\right)=\E(\mathcal{M}_{t}(\x_{t-1})+\w_t\mid \y_{1: t-1})$ is the forecast state and $\bfSigma_t^f$ is the forecast error covariance $\bfSigma_t^f=\operatorname{var}((\x_{t}-\x_t^f)\mid \y_{1: t-1})$. In fact, $\hat\x_{t}^{a}$ defined by \eqref{enkfupdate} is the solution to the minimizing problem \eqref{objectjx} 
\begin{equation} \label{objectjx}
		J(\x)=\frac{1}{2}\big\{(\x-\hat\x_t^f)^{\T}(\hat\bfSigma_t^f)^{-1}(\x-\hat\x_t^f)+\left(\y_t-\mathbf{H}_t \x\right)^{\T} \R_t^{-1}\left(\y_t-\mathbf{H}_t \x\right)\big\},
		\end{equation}
        and $\x_{t}^{a}$ defined by \eqref{analysisupdate1}-\eqref{kalmangain}  is the solution to the above minimizing problem
  with $\hat\x_t^f$ and $\hat\bfSigma_t^f$ replaced by   $\x_t^f$ and $\bfSigma_t^f$. 
For the ease of reference,  and to avoid differentiating between the linear and the non-linear cases, we also call $\x_{t}^{a}$ as the analysis state of the Kalman Filter even in the non-linear model case. 
The next two sections will establish the convergence of the EnKF analysis state  $\hat{\x}_{t}^{a}$ to the underlying $\x_{t}^{a}$ of the Kalman Filter.   
\section{Convergence of Ensemble Kalman Filter } \label{sec-enkfconvergefix}
We first provide the upper bound of the estimation error of the estimated Kalman Gain matrix.
Throughout the paper, 
$\|\cdot\|$ denotes the  $L_2$ (spectral) norm of a matrix or the $L_2$ norm of a vector,  while other matrix norms will be specified explicitly. For two positive sequences $a_{n}$ and $b_{n}$, $a_{n} \asymp b_{n}$, $a_{n} \ll b_{n}$ and $a_{n} \gg b_{n}$ mean that $a_{n} / b_{n}$ is bounded away from zero and infinity, $a_{n} / b_{n}\rightarrow 0$ and $b_{n} / a_{n}\rightarrow 0$  as $n \rightarrow \infty$, respectively.    The following assumptions are needed in the analysis. 

 	\begin{assumption}\label{ass1}
(i) The model operator $\mathcal{M}_t$ is Lipschitz continuous, that is $\|\mathcal{M}_t(\x)-\mathcal{M}_t(\x^{\prime})\|\leq \ell_0\|\x-\x^{\prime}\|$ for any $\x,\x^{\prime}\in R^p$ for a positive constant $\ell_0$.
 (ii) 
The model operator $\mathcal{M}_t$ can be locally linearized and satisfies $\|\mathcal{M}_t(\x)-\mathcal{M}_t(\x^{\prime})\|\leq \|\dot{M}_{t}\|\|\x-\x^{\prime}\|$ for $\x,\x^{\prime}\in R^p$ and $\|\x-\x^{\prime}\|\leq \|\h\|$  for a small positive $\h$,   
where $\dot{M}_{t}$ is a matrix related to the linear expansion of $\mathcal{M}_t$ at $\x$ and $\|\dot{M}_t\|\leq \ell_t$ for some positive constants $\ell_t$.
	\end{assumption} \begin{assumption}\label{ass2}
		The model error covariance $\Q_t$, the observation error covariance  $\R_t$ and the initial analysis covariance  $\bfSigma_0^a$  satisfy
		$0<\varepsilon_0\leq\lambda_{\min }\left(\Q_t\right) \leq \lambda_{\max }\left(\Q_t\right) \leq 1/\varepsilon_0$,
		$0<\varepsilon_0\leq\lambda_{\min }\left(\R_t\right) \leq \lambda_{\max }\left(\R_t\right)\leq 1/\varepsilon_0,$
and $\lambda_{\max }\left(\bfSigma_0^a\right) \leq 1/\varepsilon_0,$
for a positive $\varepsilon_0$, respectively, where $\lambda_{\max }$ and $ \lambda_{\min }$ are the maximum and minimum eigenvalue  operators, respectively.		
	\end{assumption}
	\begin{assumption}\label{ass3}
		The observation operator $\H_t$ satisfies $\|\H_t\| \leq C$ for a constant $C>0$.
	\end{assumption}
 
Assumption \ref{ass1} (i) and (ii) are two versions of the condition on the model operator $\mathcal{M}_t$. If model $\mathcal{M}_t$ is linear, it automatically satisfies both (i) and (ii) of Assumption \ref{ass1}. If $\mathcal{M}_t$ is nonlinear and Lipschitz continuous, $\mathcal{M}_t(\cdot)$ has at most linear growth at infinite intervals.
However, strongly nonlinear systems may not satisfy the Lipschitz condition. In that case, we assume  $\mathcal{M}_t$  satisfies 
 condition (ii), which is related to the tangent linear hypothesis \citep{thepaut1991four,da1999} in the 4D-Var, where $\mathcal{M}_t$ is determined in the linear approximation $
\mathcal{M}_t(\mathbf{x}_t+\mathbf{h})\approx \mathcal{M}_t(\mathbf{x}_t)+\dot{M}_t \mathbf{h}
$ in the vicinity of $\mathbf{x}_t$ for a perturbation $\mathbf{h}$, and the operator $\dot{M}_t $ 
   is called the differential or tangent linear function of $\mathcal{M}_t$ at point $\x_t$.  Assumption \ref{ass2} assumes  the eigenvalues of the three basic  covariances  $\Q_t$, 
  $\R_t$ and $\bfSigma_0^a$ are bounded away from zero and infinity.
   From Lemma  S.3.1 of the SM,  the eigenvalues of $\bfSigma_t^f$ satisfy the conditions for the thresholding and bandable(or circular bandable) covariance classes under Assumption \ref{ass2} and Assumptions \ref{ass1}-\ref{ass2}, respectively. 
 Assumption 
 \ref{ass3} puts
 a mild condition on the observation operator  $\H_t$.

The following lemma provides a basic result that bounds $\|\hat\K_t-\K_t\|$ for both fixed and high dimensional situations,  where 
$\hat\K_t=\hat\bfSigma_t^f \mathbf{H}_t^{\mathrm{T}}\big(\mathbf{H}_t \hat\bfSigma_t^f \mathbf{H}_t^{\mathrm{T}}+\mathbf{R}_t\big)^{-1}$
is an estimator of the Kalman Gain matrix assuming $\R_t$ is known. Closely related forms have appeared in the literature \citep{legland2009large, kwiatkowski2015convergence}. 

\begin{lemma}\label{kalmanconsit}
	Suppose $\R_t$ is known and
    $\lambda_{\min }\left(\R_t\right)>0$, 
    then 
   $$  \|\hat\K_t-\K_t\|\leq \lambda_{\min }^{-1}\left(\R_t\right)\|\I-\K_t\H_t\| {\left\|\H_t\right\|}\|\hat\bfSigma_t^f-\bfSigma_t^f\|
  \leq  {\lambda_{\min }^{-1}\left(\R_t\right)}{\left\|\H_t\right\|}\|\hat\bfSigma_t^f-\bfSigma_t^f\|.$$
	\end{lemma}	
 Lemma \ref{kalmanconsit}
suggests that the upper bound of the estimation error in $ \|\hat\K_t-\K_t\|$ mainly depends on the bound of $\|\hat\bfSigma_t^f-\bfSigma_t^f\|$, which 
is valid for both fixed and high dimensions.
Note that under fixed dimension, 
 the sample covariance is consistent with the underlying population covariance at the $\sqrt{n}$ rate, that is 
 $\|\hat\bfSigma_t^f-\bfSigma_t^f\|=O_P({n}^{-1/2})$ and thus $\|\hat\K_t-\K_t\|=O_P({n}^{-1/2})$. 
 The corresponding results for the high dimensions are shown later.

\begin{remark}\label{remark1}
      The last inequality in Lemma \ref{kalmanconsit} is derived from $\|\I-\K_t\H_t\| <1$ as $(\I-\K_t\H_t)\bfSigma_t^f=\bfSigma_t^a$ is the  conditional variance of $\x_t|\y_{1:t}$ and 
$(\I-\K_t\H_t)\bfSigma_t^f=\bfSigma_t^a < \bfSigma_t^f$  as long as $\x_t$ is correlated with $\y_{1:t}$. Let us take a close look at $\|\I-\K_t\H_t\|$. For the extreme case of $\R_t=\0$,  $\|\I-\K_t\H_t\|$  would be $0$ if $\H_t$ is invertible. Intuitively, $\|\I-\K_t\H_t\|$ would be close to $0$ if $\R_t$ is close to $\0$.
Suppose $p=q=1$, then $\|\I-\K_t\H_t\|=(1+\H_t^2\sigma_{\bfSigma_t^f}^2/\sigma_{\R_t}^2)^{-1}<1$. This means that  $\|\I-\K_t\H_t\|$ decreases with the increase of $\H_t^2\sigma_{\bfSigma_t^f}^2$ or decrease of $\sigma_{\R_t}^2$. For $p,q>1$, if $\H_t=\I$, then $\|\I-\K_t\H_t\|=\|\I-\bfSigma_t^f(\bfSigma_t^f+\R_t)^{-1}\|=\varepsilon_0^2/(1+\varepsilon_0^2)$ if we assume $\R_t=\varepsilon_0^2\bfSigma_t^f$. This means that  $\|\I-\K_t\H_t\|$ will decrease with  the decrease of $\varepsilon_0^2$, the ratio of $\R_t$ to $\bfSigma_t^f$.
 \end{remark}
 
In practice, the true observation error covariance matrix $\R_t$ may be unknown. 
Let $\hat\R_t$ be an estimator of $\R_t$ based on either knowledge or historical data. The following Corollary \ref{hatkasympprop}  provides a result that bounds $\|\hat\K_t-\K_t\|$ for both fixed and high dimensional situations,  where 
$\hat\K_t=\hat\bfSigma_t^f \mathbf{H}_t^{\mathrm{T}}\big(\mathbf{H}_t \hat\bfSigma_t^f \mathbf{H}_t^{\mathrm{T}}+\mathbf{\hat R}_t\big)^{-1}$
is an estimator of the Kalman Gain matrix when both the forecast and observation error covariance matrices $\bfSigma_t^f$ and $\R_t$ are estimated by $\hat\bfSigma_t^f$ and $\hat\R_t $, respectively. 
%
\begin{corollary}
\label{hatkasympprop}
If both the forecast and observation error covariance matrices $\bfSigma_t^f$ and $\R_t$ are estimated by $\hat\bfSigma_t^f$ and $\hat\R_t $, respectively, then 
  		\begin{align*}\|\hat\K_t-\K_t\|&\leq 
        \big(\|\I-\K_t\H_t\|\|\hat\bfSigma_t^f-\bfSigma_t^f\|\|\H_t\| +
\|\K_t\|\|\R_t-\hat\R_t\|\big)\|(\H_t\hat\bfSigma_t^f\H_t^{\T}+\hat\R_t)^{-1}\|.
				\end{align*}
  If $\R_t$ is invertible and $\hat\R_t$ is invertible with probability approaching $1$,
 then $$\|\hat\K_t-\K_t\|\leq {\lambda_{\min }^{-1}(\hat\R_t)}{\left\|\H_t\right\|}\|\hat\bfSigma_t^f-\bfSigma_t^f\|+\lambda_{\min }^{-1}(\hat\R_t)\lambda_{\min }^{-1}(\R_t)\left\|\H_t\right\|\|\bfSigma_t^f\|\|\R_t-\hat\R_t\|.$$
   \end{corollary}  

Compared with the upper bound in Lemma \ref{kalmanconsit}, there is an extra term related to the bound of the estimation error in the observation error covariance matrix  
 $\R_t$.
Most of the results established in the following of the paper are for the case of known 
$\R_t$ as is commonly assumed in the literature of the EnKF. Results for estimated $\R_t$ can be attained in the fashion shown in Corollary \ref{hatkasympprop}.

Based on the result in Lemma \ref{kalmanconsit} (or Corollary
\ref{hatkasympprop} for unknown $\R_t$), we can establish the convergence of the EnKF forecast and  analysis states $\hat{\x}_{t}^{f}$ and $\hat{\x}_{t}^{a}$ to their oracle Kalman Filter counterparts $\x_{t}^{f}$ and $\x_{t}^{a}$ 
based on the true $\bfSigma_t^f$ and the true underlying evolution given in 
 \eqref{analysisupdate1}-\eqref{kalmangain} by firstly establishing the convergence of 
$ \hat\bfSigma_t^f$ to $\bfSigma_t^f$. 
For the fixed  or low dimensional state vectors $\x_t$ where $ p$ is of a smaller order of the ensemble size $n$ such that $p/n \to 0$, 
 the sample covariance matrix is consistent such that
 $\|\hat\bfSigma_t^f-\bfSigma_t^f\|=O_P({n}^{-1/2})$.

 However,
remote sensing technology has been advancing the measurement of massive amounts of data sets for many geophysical processes, which produces high resolution observational data. Meanwhile, the increasing demands of high resolution forecast and dataset also 
make the spatiotemporal resolution of the state variables become much higher. Both 
 make $\x_t$ high dimensional and bring high dimensional challenges on the validity of the EnKF. 
When the dimension $p$ of the state vector $\x_t$ is  high so that  $p\gg n$, the forecast  ensemble sample covariance $\hat{\bfSigma}_t^f$ in (\ref{samplecov})  is no longer consistent to the underlying covariance $\bfSigma_t^f$
\citep{bai1993},  
which is a major threat to the validity of the EnKF.  
The last two decades have witnessed rapid development in 
high dimensional statistical inference. Consistent \hd\  covariance estimators have been proposed, which include the banding,  tapering and thresholding estimators 
 \citep{bickelRegularizedEstimationLarge2008,  bickelCovarianceRegularizationThresholding2008, caiOptimalRatesConvergence2010}.  

 Next, we present several 
consistent estimators of the high dimensional forecast error covariance  $\bfSigma_t^f$ and their covariance classes
before we evaluate the accuracy of the \hd\ assimilated state $\hat \x_t^{a}$ from the EnKF to the underlying $\x_t^a$ of the Kalman Filter. 
 \subsection{Consistent  estimation of high dimensional
 forecast error covariance matrix
 }\label{sec-4-1-matrix-estimation} 
 

The banding estimator \citep{bickelRegularizedEstimationLarge2008} of the forecast error covariance matrix with bandwidth $k $  is $$B_k(\S_{t}^f)=\big[s_{ti j}^f\mathbf{1}(|i-j| \leq k)\big]_{p \times p},\,\,0 \leq k<p,$$ where   $\S_{t}^f=[s_{ti j}^f]_{p \times p}$ is the sample covariance of the EnKF forecast ensembles. 
The  covariance matrix class suitable for the banding estimator is  
the following  
bandable covariance class, $		\mathcal{U}_b\left(\varepsilon_0, \alpha, C\right)=\big\{\bfSigma=(\sigma_{ij})_{p\times p}: \max _j \sum_{|i-j|>k}\left|\sigma_{i j}\right| \leq C k^{-\alpha} \text { for all } k>0 \text { and } 0<\varepsilon_0 \leq \lambda_{\min }(\bfSigma) \leq \lambda_{\max }(\bfSigma) \leq 1 / \varepsilon_0\big\}   $
 for some positive $\alpha$, $C$ and $\varepsilon_0$.  
 This bandable pattern is generally consistent with the dependence between variables observed at sites $i$ and $j$ in  geoscience studies, as the dependence becomes weaker as the distance between the two sites gets larger. 

The banding estimator administrates a hard threshold as it sets all matrix elements
beyond the $k$-th  sub-diagonals to zero. 
 \cite{caiOptimalRatesConvergence2010} updated it with a tapering alteration which tapers
down the level of the thresholding gradually
	 $$	 
  \hat{\bfSigma}_{t,tap}^f(k)=\S_{t}^f \circ \W(k)   
  = \big[w_{i j}(k)  s_{t i j}^f \big]_{p \times p},
	 $$
 where 
 $\W(k) = (w_{ij}(k))_{p \times p}$, with  
	 $
	 w_{i j}(k)=(2/k) \big\{(k-|i-j|)_{+}-(k/2-|i-j|)_{+}\big\}, 
	 $ 
is the tapering matrix at a tapering width $k$ which is usually an even integer 
	 and $1 \leq k \leq p$. 

\begin{figure}[!ht]
	\centering
	\subfigure[{\scriptsize $h(k)$ of $ \bfSigma \in \mathcal{U}_b$}]{
		\includegraphics[width=0.35\textwidth]{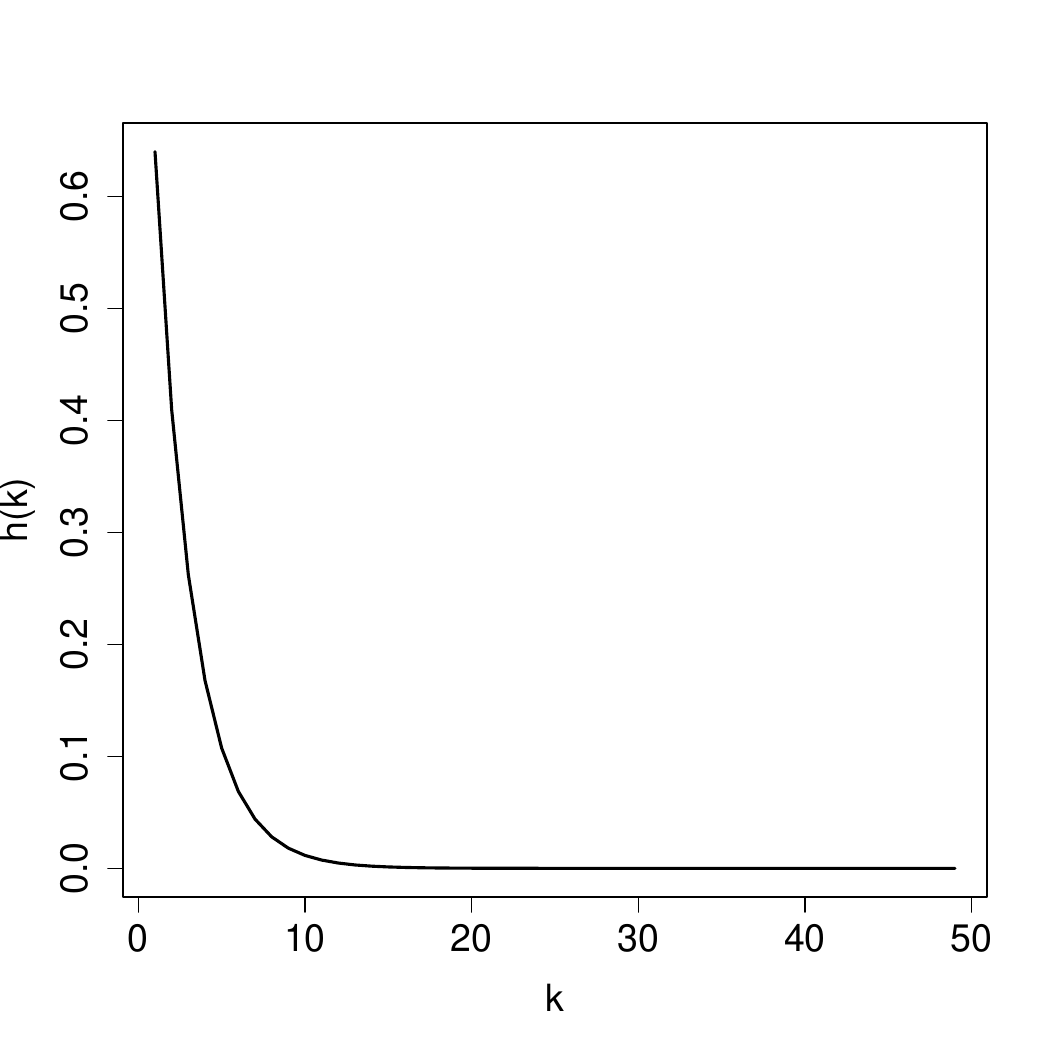}}
	\subfigure[{\scriptsize $h(k)$ of $ \bfSigma \in \mathcal{U}_{bc}$}]{
		\includegraphics[width=0.35\textwidth]{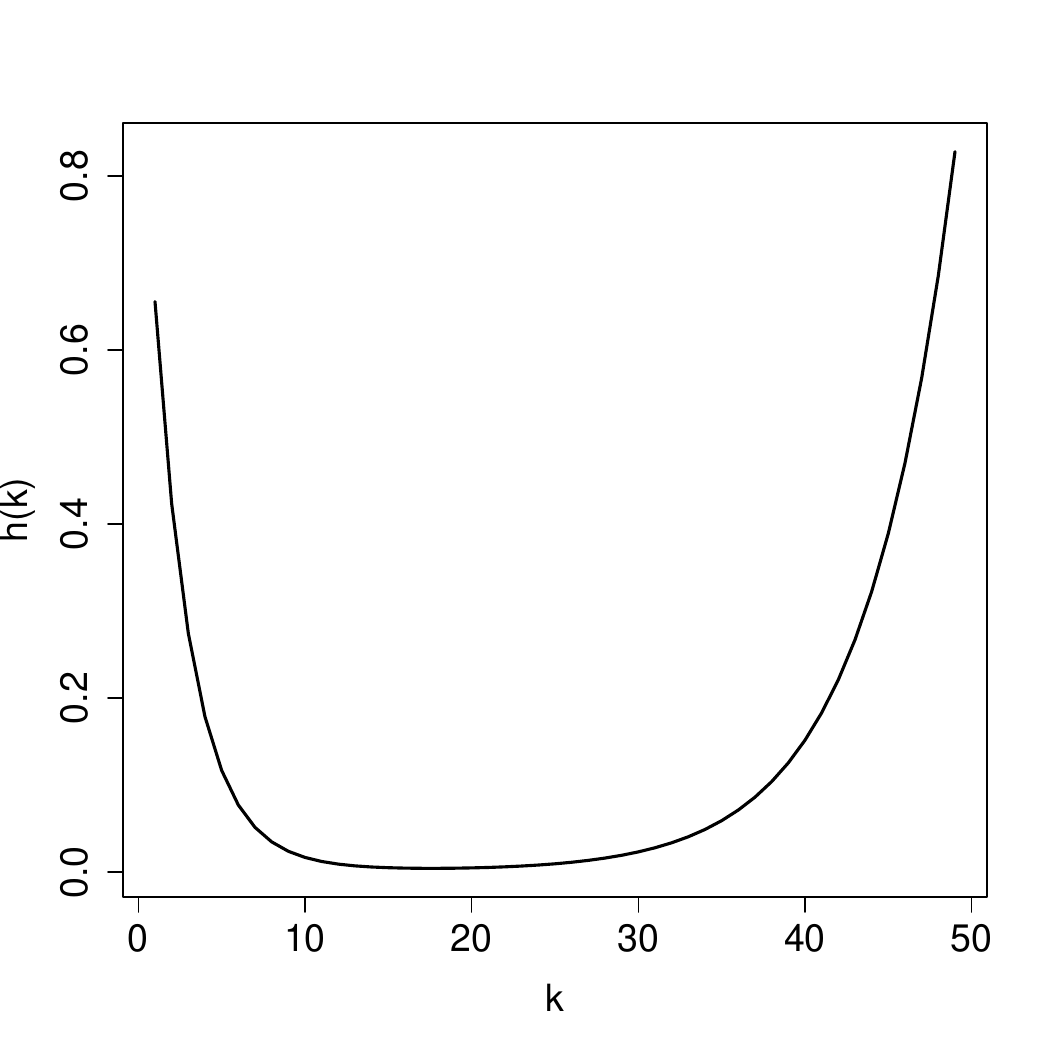}}\\
	\subfigure[{\scriptsize Heatmap of $ \bfSigma \in \mathcal{U}_b$}]{
		\includegraphics[width=0.35\textwidth]{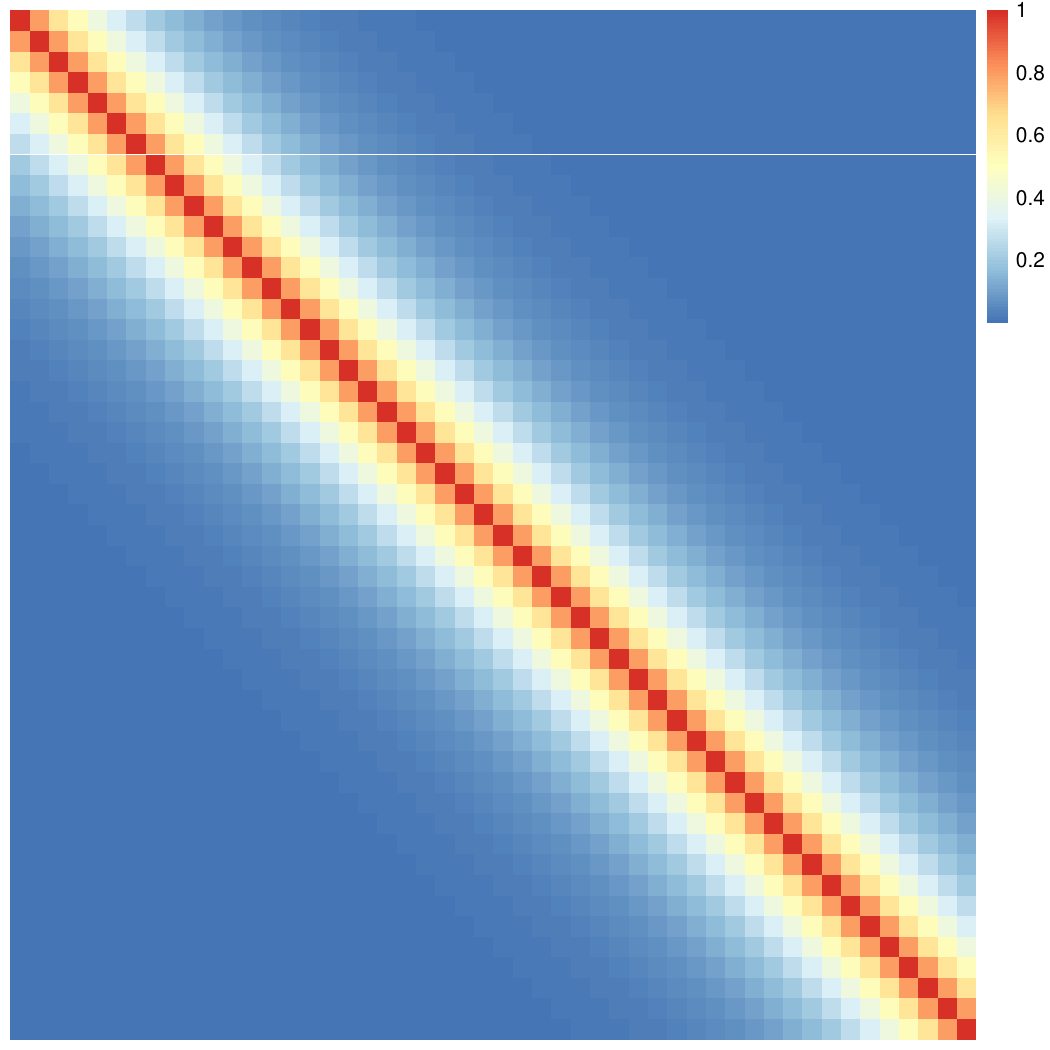}}
	\subfigure[{\scriptsize Heatmap of $ \bfSigma \in \mathcal{U}_{bc}$}]{
		\includegraphics[width=0.35\textwidth]{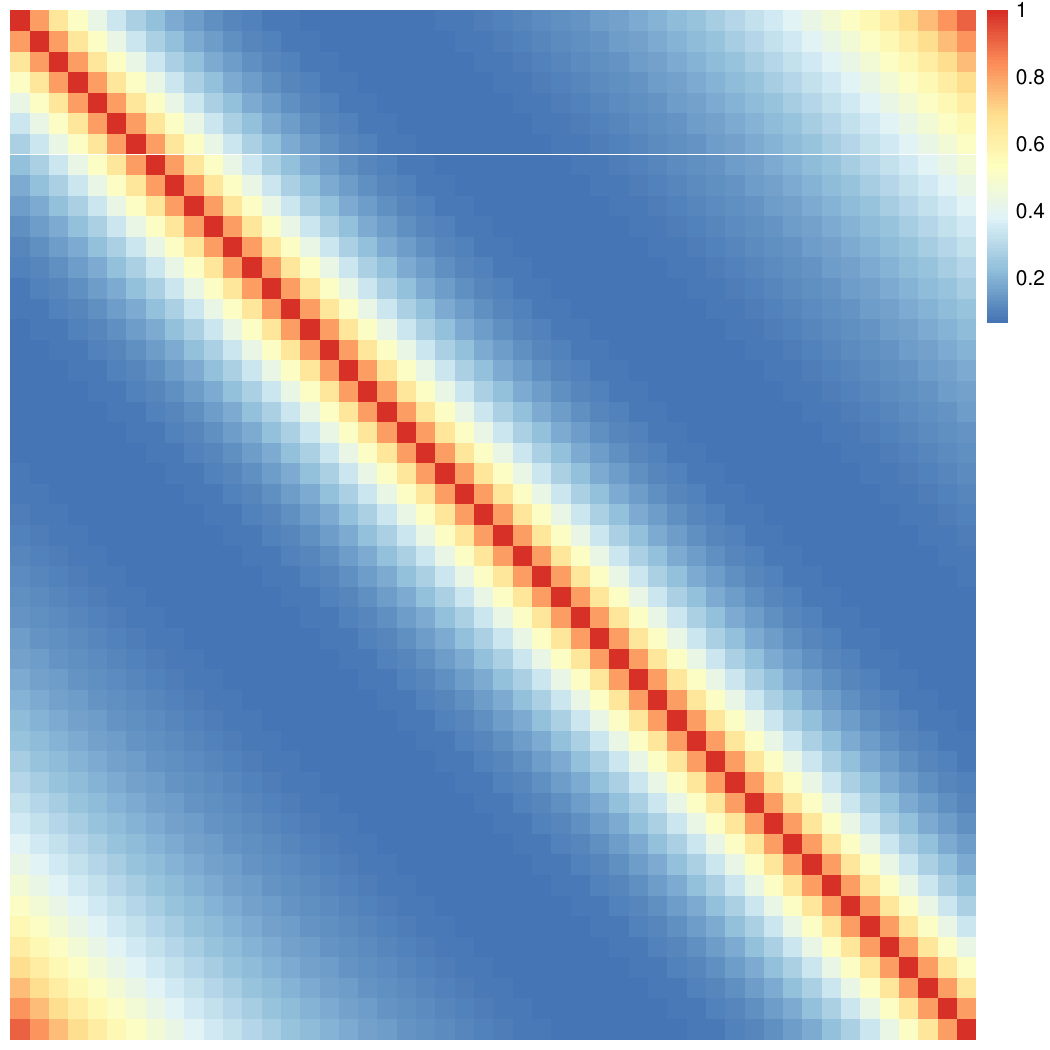}}
		\caption{Bandable $\mathcal{U}_b$ and circular bandable $\mathcal{U}_{bc}$ covariance matrices: the average energy function   $h(k)$ in Panels (a) and (b) and their  heatmaps in Panels (c) and (d), 
  where $h(k)$ is the average of the squares of the $k^{th}$ sub-diagonal entries.}
	\label{bandingar}
\end{figure} 
In earth science, due to the largely spherical shape of the earth, for a study region that encompasses the entire earth or  its poles,  
the state vectors are highly dependent for the variable components between the two ends of the state vector.  
Thus we propose a circular form of the bandable structure, $	\mathcal{U}_{bc}\left(\varepsilon_0, \alpha, C\right)=\big\{\bfSigma=(\sigma_{ij})_{p\times p}: \max _j \sum_{k_1<|i-j|<p-k_2}\left|\sigma_{i j}\right| \leq C (k_1+k_2)^{-\alpha} \text { for all } 
		0<k_1,k_2<p\text { and } 0<\varepsilon_0 \leq \lambda_{\min }(\bfSigma) \leq \lambda_{\max }(\bfSigma) \leq 1 / \varepsilon_0\big\}.$
Figure \ref{bandingar} depicts the $h(k)$ functions and heatmaps of two specific covariances that belong to the two bandable covariance classes $\mathcal{U}_b$ and $\mathcal{U}_{bc}$, where $h(k)$ 
is the average of the squares of the $k^{th}$ subdiagonal entries of a covariance defined by
$h(k):=0.5{(p-k)^{-1}} \sum_{\left|l_1-l_2\right|=k} \sigma_{l_1 l_2}^2=(p-k)^{-1}\sum_{l=1}^{p-k} \sigma_{l,l+k}^2.$ Figures \ref{bandingar} (a)-(b) show that the $h(k)$ functions of covariances in $\mathcal{U}_b$ and $\mathcal{U}_{bc}$ are $L$-type and $U$-type curves as the subdiagonal index $k$ increases, respectively.
For covariances  in
    the circular bandable class $\mathcal{U}_{bc}\left(\varepsilon_0, \alpha, C\right)$, the proposed mid-banding forecast error covariance estimator with bandwidths $k_1$ and $k_2$ 
is  $$B_{k_1,k_2}(\S_{t}^f)=[s_{t i j}^f\mathbf{1}(|i-j| \leq k_1\, \text{or}\, |i-j|\geq(p-k_2))]_{p \times p},\,\,0 \leq k_1< (p-k_2)\leq p.$$

The bandable matrix class requires that the state vector  $\x_t$ has an ordering (or a permutation) such that the correlation decays as two components of $\x_t$ are further apart. 
However, the presence of geo-spatial processes  (weather front or hurricane) and geography (mountains or ocean)   will interrupt the spatial dependence  ordering prescribed by the banding covariance class. In such situations, the  thresholding covariance  class $\mathcal{U}_\tau\left(\gamma, c_0(p), M, \varepsilon_0\right)$ 
suitable for the thresholding estimator \citep{bickelCovarianceRegularizationThresholding2008}  may be employed, where 
$ \mathcal{U}_\tau\left(\gamma, c_0(p), M, \varepsilon_0\right)=\big\{\bfSigma=(\sigma_{ij})_{p\times p}: \sigma_{i i} \leq M, \sum_{j=1}^p\left|\sigma_{i j}\right|^{\gamma} \leq c_0(p), \text { for all } i \text{ and } \lambda_{\min }(\bfSigma) \geq \varepsilon_0>0\big\}, \text{ for } 0 \leq \gamma<1.$
The thresholding forecast error covariance matrix estimator with a threshold level $s$ is 
 $T_{s}({\S}_t^f)=\big[s_{ti j}^f\mathbf{1}(|s_{ti j}^f|  \geq s)\big]_{p \times p}$. 

Note that all of the banding, mid-banding, tapering and thresholding estimators depend on either the bandwidth $k$ or the threshold $s$ which can be 
selected by methods in \cite{qiuBandwidthSelectionHighDimensional2015}. For the bandwidths $k_1,k_2$ of the mid-banding estimator,
 we propose a similar selection method 
and details can be found in Section S.2 of the SM.

 To apply the 
 \hd\ estimators of the forecasting error covariance  $\bfSigma_t^f$, 
  we need the following distributional assumption.  
\begin{assumption}\label{ass_band}
The random vectors $\mathbf{x}_{t,i}^f=(x_{t,i1}^f,\cdots,x_{t,ip}^f)^{\T}$ are IID and  each $x_{t,1j}^f\stackrel{d}{\sim} F_{t,j}$ and $(x_{t,1j}^f)^2\stackrel{d}{\sim}G_{t,j}$, where $G_{t,j}(u)=F_{t,j}(\sqrt u)-F_{t,j}(-\sqrt u)$ for $u>0$. 
There exists a $\lambda_0>0$ such that 
$\max_{1\leq j\leq p}\E(e^{\lambda (x_{t,1j}^f)^2})<\infty\textrm{ for }\lambda\in (-\lambda_0,\lambda_0).$
\end{assumption}
Assumption \ref{ass_band}  
effectively assume $(x_{t,1j}^f)^2$ are sub-exponentially distributed, which include the Sub-Gaussian random variables as a special case.

The following Proposition \ref{converge_of_sigmatf} 
summarizes the convergence rates of the forecast error covariance estimators by the banding, mid-banding,  tapering and thresholding methods. 
\begin{proposition}\label{converge_of_sigmatf}   Suppose the  forecast ensemble $\{\mathbf{x}_{ti}^f\}_{i=1}^n$ are 
conditionally
sub-exponentially distributed as in Assumption \ref{ass_band}. 
\begin{itemize}
 \item 
   [(i)] For the banding estimator $B_k(\S_{t}^f)$ with $k \asymp (n^{-1} \log p )^{-1 /(2\alpha+2)}$, then  uniformly for $\bfSigma_t^f \in \mathcal{U}_b\left(\varepsilon_0, \alpha, C\right)$,
   $\E\|B_k(\S_{t}^f)-\bfSigma_t^f\|^2\asymp (n^{-1}{\log p})^{\alpha /(\alpha+1)}.$ 
    \item [(ii)] For the mid-banding estimator $B_{k_1,k_2}(\S_{t}^f)$ 
    with $k_1+k_2 \asymp(n^{-1} \log p)^{-1/ (2\alpha+2)}$, uniformly on $\bfSigma_t^f \in \mathcal{U}_{bc}\left(\varepsilon_0, \alpha, C\right)$, $\E\|B_{k_1,k_2}(\S_{t}^f)-\bfSigma_t^f\|^2  \asymp (n^{-1}{\log p})^{\alpha /(\alpha+1)}.$
  \item[(iii)] For the tapering estimator $\hat{\bfSigma}_{t,tap}^f(k)$ 
and  $\bfSigma_t^f \in \mathcal{U}_b(\varepsilon_0,\alpha,C)$,   the optimal rate of convergence of the tapering estimator is $	\mathbb{E}\|\hat{\bfSigma}_{t,tap}^f(k)-\bfSigma_t^f\|^2 \asymp \min \big\{n^{-2 \alpha /(2 \alpha+1)}+{n}^{-1}{\log p}, \;{n}^{-1}{p}\big\}.
		$
    \item[(iv)] For the thresholding estimator $T_{s}({\S}_t^f)$, 
    for sufficiently large $M^{\prime}$, if $s=M^{\prime} (n^{-1}{\log p})^{-1/2}$, then, uniformly on $\mathcal{U}_\tau\left(\gamma, c_0(p), M\right)$,  $
		\E\|T_{s}({\S}_t^f) -\bfSigma_t^f\|^2  \asymp c_0^2(p)(n^{-1}{\log p})^{(1-\gamma)}.
  $
  %
  \end{itemize}
 
  \end{proposition}
The results in  Proposition \ref{converge_of_sigmatf} (i), (iii) and (iv) are, respectively, from \cite{bickelRegularizedEstimationLarge2008}, \cite{caiOptimalRatesConvergence2010} and \cite{bickelCovarianceRegularizationThresholding2008}.  And that for the mid-banding estimator in   Proposition \ref{converge_of_sigmatf} (ii) is a new result whose proof is given in the SM.  
 They imply that 
 the \hd\ covariance estimators are  
consistent to $\bfSigma_t^f$ under the spectral norm if $\log(p) =o(n)$ or $p =o(e^{c n})$ for any positive constant $c$.  

According to Lemma \ref{kalmanconsit}, the convergence results in Proposition \ref{converge_of_sigmatf} on the \hd\ estimation of  $\bfSigma_t^f$ translate to that of the estimated Kalman Gain $\hat\K_t=\hat\bfSigma_t^f \mathbf{H}_t^{\mathrm{T}}(\mathbf{H}_t \hat\bfSigma_t^f \mathbf{H}_t^{\mathrm{T}}+\mathbf{R}_t)^{-1}$  to  $\K_t$.  
Specifically,  
    \begin{align}
  & 
  \text{ if }\bfSigma_t^f \in \mathcal{U}_b\left(\varepsilon_0, \alpha, C\right)\text{or } 
\mathcal{U}_{bc}\left(\varepsilon_0, \alpha, C\right), 
  \E\|\hat\K_t-\K_t\|^2 \asymp (\frac{\log p}{n})^{\frac{\alpha}{\alpha+1}}\, \text{ for } \hat \bfSigma_t^f=B_k(\S_{t}^f)\text{ or }B_{k_1,k_2}(\S_{t}^f);\nn
  \\
&\text{ if }\bfSigma_t^f \in \mathcal{U}_b\left(\varepsilon_0, \alpha, C\right), 
\E\|\hat\K_t-\K_t\|^2 \asymp \min \big\{n^{\frac{-2\alpha}{2\alpha+1}}+\frac{\log p}{n}, \;\frac{p}{n}\big\}   \text{ for }\, \hat \bfSigma_t^f=\hat{\bfSigma}_{t,tap}^f(k);  \nn
\\
& \text{ if }\bfSigma_t^f \in \mathcal{U}_\tau\left(\gamma, c_0(p), M\right),  
 	\E\|\hat\K_t-\K_t\|^2 \asymp c_0^2(p)(\frac{\log p}{n})^{(1-\gamma)} \ \text{ for } \hat \bfSigma_t^f=T_{s}(\S_{t}^f).\label{eq:kt-banding} 
   \end{align}

 As will be shown later, the Kalman Gain estimation error $\Delta_{\K,t}=\E\|\hat\K_t-\K_t\|^2$ plays a crucial role in quantifying the effects of the \hd\  banding, mid-banding, tapering and  thresholding 
 estimators for $\bfSigma_t^f$. 
	
\subsection{Convergence of High Dimensional Ensemble Kalman Filter}\label{sec-enkfconvergehd} 
When the dimension $p$ of $\x_t$ is  high,  
to allow consistency of the data assimilation,  we have  to use the \hd\ 
covariance estimators  in Section \ref{sec-4-1-matrix-estimation} for the forecast error covariance matrix $\bfSigma_t^f$ and the Kalman Gain matrix $\K_t$. 
Based on the consistent estimators of $\bfSigma_t^f$ and  $\K_t$, we propose the HD-EnKF algorithms given in Section S.1  of the SM and we establish the accuracy of the assimilated state $\hat \x_t^{a}$ to the underlying $\x_t^a$ in the following. 
To simplify notation, we denote 
 		 $\Delta_{\K,k}=\mathbb{E}\|\hat\K_k-\K_k\|^2$  
 and 
  \begin{align*}
 \Omega_{t}= 
 &\operatorname{tr}\big({n}^{-1}\E(\hat\K_t-\K_t)^{\T}(\hat\K_t-\K_t)({{\H}_t \bfSigma_{t}^f{\H}_t^{\T}+\R_t})\big) +\E\|(\hat\K_t-\K_t)(\y_t-{\H}_t \x_{t}^f)\|^2\nn
\\&-2\operatorname{tr}\big((\I-\K_t\H_t)^{\T}\E(\hat\K_t-\K_t)\H_t \bfSigma_{t}^f/n\big)+2\operatorname{tr}\big(\big(\K_t^{\T}\E(\hat\K_t-\K_t)\R_t/n\big)\big).
 \end{align*}
The following Theorem \ref{enkfconsit_1stephigh}  presents the upper bounds of the discrepancy between the EnKF analysis state and the underlying Kalman Filter analysis for the one-step data assimilation for a general case where the model operator $\mathcal{M}_t$ is possibly nonlinear.
	\begin{theorem}	\label{enkfconsit_1stephigh}
		 Suppose Assumptions  \ref{ass1}(i) or \ref{ass1}(ii),  \ref{ass2}, 
   \ref{ass3} hold
  and 
  the forecast ensemble $\{\x_{t,j}^f \}_{j=1}^n$ are  independently  sampled from the  distribution of $\x_{t} \mid \y_{1: t-1}$ with mean $\x_{t}^{f}$ and covariance $\bfSigma_{t}^f$.
   Then, 
   the forecast and analysis states of the EnKF at time $t$ satisfy
			\begin{align}
			   	&p^{-1}\mathbb{E}\{\|\hat\x_{t}^{f}-\x_t^f\|^2|\y_{1:t-1}\}={n}^{-1}p^{-1}\operatorname{tr}(\bfSigma_{t}^f)=O\big(n^{-1}\big),\nn
   \\
				&p^{-1}\mathbb{E}\{\|\hat\x_{t}^{a}-\x_t^a\|^2|\y_{1:t}\}= \operatorname{tr}\big(n^{-1}p^{-1}(\I-\K_t\H_t)\bfSigma_{t}^f\big)+p^{-1}\Omega_{t}\label{eq:mseaonestephigh}
		\\&\quad\quad\quad\quad\quad\quad\quad\quad\quad\;\,= O\big(n^{-1}+p^{-1}\Delta_{\K,t}\|\y_t-{\H}_t \x_{t}^f\|^2\big).\nn	 
			\end{align}
	\end{theorem}

Theorem \ref{enkfconsit_1stephigh} gives the upper bounds of the one-step data assimilation MSE with high dimensional forecast error covariance estimators, which depends on $\Delta_{\K,t}$ whose convergence rates for different covariance estimators are given in \eqref{eq:kt-banding}.      Note that the first term in \eqref{eq:mseaonestephigh}, $n^{-1}(\I-\K_t\H_t)\bfSigma_{t}^f=n^{-1}\bfSigma_{t}^a$,  is the variance of the sample mean of the analysis ensemble using the true forecast error covariance $\bfSigma_t^f$  and the second term $\Omega_{t}$ results from the estimation error of $\hat\bfSigma_t^f$. 
Specifically, with the banding or  mid-banding estimators, 
the tapering estimator and the thresholding estimator, 
the one-step EnKF data assimilation MSE are, respectively, 
   \begin{align}
   \label{eq:hd1stepbanding}
 p^{-1}\mathbb{E}\{\|\hat\x_{t}^{a}-\x_t^a\|^2|\y_{1:t}\}&\asymp \max\{n^{-1},\; p^{-1}q(n^{-1}{\log p})^{\alpha /(\alpha+1)}\}, \\
      p^{-1}\mathbb{E}\{\|\hat\x_{t}^{a}-\x_t^a\|^2|\y_{1:t}\}&\asymp \max\big\{n^{-1},\; p^{-1}q(n^{\frac{-2 \alpha}{2 \alpha+1} }+{n}^{-1}{\log p})\big\},\label{eq:hd1steptapering}\\
          p^{-1}\mathbb{E}\{\|\hat\x_{t}^{a}-\x_t^a\|^2|\y_{1:t}\}&\asymp \max\big\{n^{-1},\; p^{-1}qc_0^2(p)(n^{-1}{\log p})^{(1-\gamma)}\big\}\label{eq:hd1stepthres}. 
  \end{align}
It should also be noted that the above results are valid for general model operators $\mathcal{M}_t$ given in Assumption \ref{ass1}-(i) and (ii).
Compared with the results in \cite{alghattas2024non}, they investigated the difference between the ensembles of single-step Ensemble Kalman Inversion using the localized estimation and the true forecast error covariance matrices, under the framework of the effective dimension of a matrix $\A$ defined by $\operatorname{tr}(\A)/\|\A\|$ (or the max-log 
 effective dimension for a sparse covariance). 


For fixed $p,q$, both the forecast and analysis states of the EnKF converge to the underlying forecast and analysis states of the Kalman Filter with a convergence rate of $\sqrt{n}$ as $n\rightarrow \infty$ because $\Delta_{\K,t}=O(n^{-1})$, as also shown in Theorem  S.3.1  of the SM. Furthermore, for linear model operator ${\M}_t$, Proposition  S.3.1 of the SM implies that with Gaussian initial state, the asymptotic conditional distributions of the analysis ensemble $\x_{t,j}^a|\y_{1:t}$ and the forecast ensemble $\x_{t,j}^f|\y_{1:t-1}$  in the EnKF are the same as those of $\x_t|\y_{1:t}$ and $\x_t|\y_{1:t-1}$ in the Kalman Filter, respectively.
  
 To show the impact of the dimensionality on the assimilation MSE, we take the banding and mid-banding estimators as examples. If $q\asymp p$,  $p^{-1}\mathbb{E}\{\|\hat\x_{t}^{a}-\x_t^a\|^2|\y_{1:t}\}	
 	=O( (n^{-1}{\log p})^{2\alpha /(2\alpha+2)}) $, which implies that the one-step assimilation MSE  mainly depends on the estimation error of the Kalman Gain.  
  Commonly in high resolution data assimilation, the dimension of the observation $\y_t$ is much smaller than that of the state vector $\x_t$, so that $q \ll p$.  
   If $q$ is relatively large such that $p^{-1}q(n^{-1}{\log p})^{\alpha /(\alpha+1)}\} \gg n^{-1}$, then the estimation error of the Kalman  Gain matrix will dominate, that is, $p^{-1}\mathbb{E}\{\|\hat\x_{t}^{a}-\x_t^a\|^2|\y_{1:t}\}	
 	=O\big(p^{-1}q(n^{-1}{\log p})^{\alpha /(\alpha+1)}\big) $. Otherwise, the first term $n^{-1}$ dominates.
   However, if sample covariance $\S_t^f$ is used in $\hat\K_t$ when $p\geq n$, then $p^{-1}\mathbb{E}\{\|\hat\x_{t}^{a}-\x_t^a\|^2|\y_{1:t}\}=O(p/n^2\|\y_t-{\H}_t \x_{t}^f\|^2)=O(pq/n^2)$. 
For the impact of the observation dimension $q$ on the MSE, a study in the SM shows that if the additional observation $y_{t,q+1}$ is not far away from its forecast ${\H}_t \x_{t}^f$, the MSE will decrease
since the first term in \eqref{eq:mseaonestephigh}  decreases more.


For the discrepancies between  the EnKF analysis state and the true state, as $\mathbb{E}\{\|\hat\x_{t}^{a}-\x_t\|^2|\y_{1:t}\}=\mathbb{E}\{\|\hat\x_{t}^{a}-\x_t^a\|^2|\y_{1:t}\}+\mathbb{E}\{\|\x_{t}^{a}-\x_t\|^2|\y_{1:t}\}$
and   $\mathbf{x}_t$ is a random variable with 
$\mathbb{E}\{\|\x_{t}^{a}-\x_t\|^2|\y_{1:t}\}= \operatorname{tr}(\bfSigma_t^a)$ which is a constant
not related to the ensemble size $n$ for a fixed $p$, then 
$\mathbb{E}\{\|\hat\x_{t}^{a}-\x_t\|^2|\y_{1:t}\}$  can not converge to zero.  This is due to the fact that the underlying state is a random variable.   In contrast, $\x_t^a$ is a conditional mean and is possible to be estimated consistently.

In practice, the ensemble forecast-analysis process in the EnKF repeats for multiple time steps ahead to make longer-time predictions. Thus, in addition to the one-step ahead data assimilation accuracy, one may also care about the
quality of multiple-step forecasting and analysis. 
The following  Theorem \ref{enkfconsit_tstephigh} gives the $h$-step MSE bounds between the forecast/analysis states of the EnKF in comparison to the underlying  Kalman Filters.  To simplify writing, 
define two non-stochastic constants 
   \begin{equation} \label{constant_CA}
   C_{t,h}^A=\frac{1-L_{t,h}^{h}A_{t,h}^{h}}{1-L_{t,h}A_{t,h}}
   \, \text{ and  }  
    C_{t,h}^B=\frac{1-L_{t,h}^{h}B_{t,h}^{h}}{1-L_{t,h}B_{t,h}},  
   \end{equation}
   where $A_{t,h}=\max_{t < k \leq t+h}\|\I-\K_{k} \H_{k}\|$, $ B_{t,h}=\max_{t < k \leq t+h}\{\E\|\I-\hat\K_{k} \H_{k}\|\}$,
  and $L_{t,h}=\max_{0< k \leq t+h}\{\ell_0$,\\$\ell_{k}\}$
 with $\ell_0$ and $\ell_k$ being respectively the Lipschitz coefficient and the upper bound of the local linearization coefficient of the model operation defined in Assumption \ref{ass1}. 
   We further denote 
\begin{align*}
U_{k}
=\big\{\| \y_{k}-{\H}_{k} \x_{k}^f\|^2\Delta_{\K,k} +\operatorname{tr}\big(\E(\hat\K_k-\K_k)^{\T}(\hat\K_k-\K_k)({\H}_{k} \bfSigma_{k}^f{\H}_{k}^{\T}+\R_{k})\big)\big\}^{{1}/{2}}. 
 \end{align*} 
 \begin{theorem}	\label{enkfconsit_tstephigh}
		Assumptions \ref{ass1}(i) or \ref{ass1}(ii),  \ref{ass2}, 
   \ref{ass3} hold
  and the analysis ensemble $\{\x_{t,j}^a \}_{j=1}^n$  is generated independently from 
   the  distribution of $\x_{t} \mid \y_{1: t}$ with mean $\x_{t}^{a}$ and covariance $\bfSigma_{t}^a$ at  time $t$.
  By repeatedly performing the ensemble forecast-analysis  process $h$ steps, the forecast and analysis states of the EnKF at the time $t+h$ satisfy
    \begin{align}
	& p^{-1}\mathbb{E}\{\|\hat\x_{t+h}^{f}-\x_{t+h}^{f}\|^2|\y_{1:t+h-1} \}
  \le  p^{-1}\Big(\{\operatorname{tr}(n^{-1}\bfSigma_{t+h}^f)\}^{1/2}
			+L_{t,h}C_{t,h-1}^{B}\max_{t< k\leq t+h-1}U_{k}\Big)^2  \nn \\  &
				= O\Big( n^{-1}
			+p^{-1}L_{t,h}^2(C_{t,h-1}^{A})^2\max_{t< k\leq t+h-1}\big\{\Delta_{\K,k}\|\y_{k}-{\H}_{k} \x_{k}^f\|^2\big\}\Big),  \nn \\
				&p^{-1}\mathbb{E}\|(\hat\x_{t+h}^{a}-\x_{t+h}^{a})|\y_{1:t+h}\|^2	
    \leq  p^{-1}\Big(\{\operatorname{tr}(n^{-1}\bfSigma_{t+h}^a)\}^{1/2}
			+C_{t,h}^{B}\max_{t< k\leq t+h}U_{k}\Big)^2 \nn \\ 
   	&= O\Big( n^{-1}
			+p^{-1}(C_{t,h}^{A})^2\max_{t< k\leq t+h}\big\{\Delta_{\K,k}\|\y_{k}-{\H}_{k} \x_{k}^f\|^2\big\}\Big). \nn 
    \end{align}
	\end{theorem}	
Compared with the one-step results in Theorem \ref{enkfconsit_1stephigh},  the term related to the estimation error in $\hat\bfSigma_{t+k}^f$ in the $h$-step analysis MSE is inflated with constants $(C_{t,h}^{A})^2$ which are related to $L_{t,h}$ and $A_{t,h}$ defined after \eqref{constant_CA}. This is because of the accumulation of the estimation error of $\hat\bfSigma_{t+k}^f$ in the data assimilation, which can be seen from the proof of Theorem \ref{enkfconsit_tstephigh} in the SM. Using a similar analysis on the one-step results after Theorem \ref{enkfconsit_1stephigh} and according to 
  \eqref{eq:kt-banding},  the $h$-step assimilation MSE between the analysis state of the EnKF $\hat\x_{t+h}^{a}$ and the  $\x_{t+h}^{a}$ for $\bfSigma_t^f$ estimated by the four \hd\ estimators are,   
 respectively, 
 \begin{align}
 \label{eq:hdhstepbanding}
    p^{-1}\mathbb{E}\|(\hat\x_{t+h}^{a}-\x_{t+h}^{a})|\y_{1:t+h}\|^2&\asymp \max\big\{n^{-1},\; (C_{t,h}^{A})^2{p}^{-1}q\big({n}^{-1}{\log p}\big)^{\alpha /(\alpha+1)}\big\},\\
      p^{-1}\mathbb{E}\|(\hat\x_{t+h}^{a}-\x_{t+h}^{a})|\y_{1:t+h}\|^2&\asymp \max\big\{n^{-1},\; (C_{t,h}^{A})^2{p}^{-1}q(n^{-\frac{2\alpha }{2\alpha+1}}+{n}^{-1}{\log p})\big\},\label{eq:hdhsteptapering}\\
          p^{-1}\mathbb{E}\|(\hat\x_{t+h}^{a}-\x_{t+h}^{a})|\y_{1:t+h}\|^2&\asymp \max\big\{n^{-1},\; (C_{t,h}^{A})^2{p}^{-1}qc_0^2(p)\big(n^{-1}{\log p}\big)^{(1-\gamma)}\big\}\label{eq:hdhstepthres}, 
  \end{align}
 in which the estimation error of $\K_{k}$ are  inflated by 
  $(C_{t,h}^{A})^2$ 
  compared with that in the upper bound of 
 the one-step assimilation MSE  given in \eqref{eq:hd1stepbanding} - \eqref{eq:hd1stepthres} in Theorem \ref{enkfconsit_1stephigh}. 
According to Remark \ref{remark1}, $A_{t,h}=\max_{t< k \leq t+h}{\left\|\I-\K_k \H_k\right\|}<1$. If $L_{t,h}A_{t,h}<1$, then  $C_{t,h}^{A}$ converges to a small constant and  $L_{t,h}^{h}A_{t,h}^{h}\rightarrow 0$ 
  as $h\rightarrow \infty$, although $C_{t,h}^{A}$ becomes larger along with the increase of $h$.
  Thus, $
\mathbb{E}\{\|\hat\x_{t+h}^{a}-\x_{t+h}^{a}\|^2|\y_{1:t+h} \}\asymp \mathbb{E}\{\|\hat\x_{t}^{a}-\x_t^a\|^2|\y_{1:t}\}$.
  If $L_{t,h}A_{t,h}>1$, then $C_{t,h}^{A}$ increases a lot with the increase of  $h$,   which suggests the quality of the data assimilation deteriorates as $h$ increases, although 
  $
\mathbb{E}\{\|\hat\x_{t+h}^{a}-\x_{t+h}^{a}\|^2|\y_{1:t+h} \}\asymp \mathbb{E}\{\|\hat\x_{t}^{a}-\x_t^a\|^2|\y_{1:t}\}$  for a fixed  $h$. 
However, it is possible to make $L_{t,h}$ be smaller by shortening the forecast time step or the integration time step.
This underscores the necessity of the high-quality or high dimensional estimators of the forecast error covariance, considering the accumulation of estimation errors in multiple-step data assimilation.
The MSE between the forecast state of the EnKF $\hat\x_{t+h}^{f}$ and the Kalman Filter $\x_{t+h}^{f}$  for the banding, mid-banding,  tapering and thresholding estimators of $\bfSigma_t^f$ are similar to \eqref{eq:hdhstepbanding}-\eqref{eq:hdhstepthres} but with slightly different inflation constants.



Sometimes, the assimilation process proceeds every $k$ time steps not every step as one may not have observations at each step or the assimilation in every step is too computationally consuming. 
Theorem  S.3.2 of the SM gives the MSE bounds for multiple-step forecasting and assimilation where the assimilation process proceeds every $k$ time steps.  

\section{Imperfect Models}\label{sec-5-imperfect-model} 
 
 In practice, the exact form of the model  operator 
  $\mathcal{M}_t$ is likely unknown, and the state-of-the-art model $\mathcal{\hat M}_t$  is only an  approximation to $\mathcal{M}_t$. Besides, large dynamic models in earth science are often nonlinear,  one may use its linear approximation \citep{julier1997new,stroudEnsembleKalmanFilter2010}.  
 This section is devoted to quantifying the effects of the model 
approximation error on the accuracy of 
the analysis and forecast states of the EnKF. 
The following assumption regulates the extent of the model approximation error. 

\begin{assumption}\label{ass4} Let  $\xi_t=\mathcal{\hat M}_t(\x)-\mathcal{ M}_t(\x)$ be the error of the approximated model $\mathcal{\hat M}_t$ to $ \mathcal{M}_t$ with  
  $\E(\xi_{t})=\mu_{\xi_t}$ and $\operatorname{Var}(\xi_{t})=\Xi_{\xi_t}$.  It is assumed that $\|\mu_{\xi_t}\|_{\infty} \leq C_1$ and $\|\Xi_{\xi_t}\| \leq C_2$ for some positive constants $C_1$ and $C_2$.  
\end{assumption}
Theorem \ref{enkfconsitM_1step} gives the one-step assimilation
 bounds of the difference between the forecast/analysis states of the EnKF and the Kalman Filter when  $\mathcal{M}_t$ is unknown.  Denote
\begin{align}
 V_{t}&= 
     \operatorname{tr}\big(n^{-1}(\I-\K_t\H_t)^{\T}(\I-\K_t\H_t)\Xi_{\xi_t}\big)+\|(\I-\K_t\H_t)\mu_{\xi_t}\|^2
     \nn \\&+\operatorname{tr}\big(n^{-1}\E(\hat\K_t-\K_t)^{\T}(\hat\K_t-\K_t)({\H}_t (\bfSigma_{t}^f+\Xi_{\xi_t}){\H}_t^{\T}+\R_t)\big)\label{eq:vt}\\&+\E\|(\hat\K_t-\K_t)(\y_t-{\H}_t \x_{t}^f-{\H}_t \mu_{\xi_t})\|^2+2\operatorname{tr}\big(\K_t^{\T}\E(\hat\K_t-\K_t)\R_t/n\big)
\nn\\& -2\operatorname{tr}\big((\I-\K_t\H_t)^{\T}\E(\hat\K_t-\K_t) \{\H_t
(\bfSigma_{t}^f+\Xi_{\xi_t})/n+(\y_t-{\H}_t \x_{t}^f-{\H}_t\mu_{\xi_t})\mu_{\xi_t}^{\T}\}\big).\nn
\end{align}

\begin{theorem}	\label{enkfconsitM_1step}
		Suppose Assumptions \ref{ass1}(i) or \ref{ass1}(ii),  \ref{ass2}, 
   \ref{ass3}, \ref{ass4} hold
  and  $\{\x_{t-1,j}^a \}_{j=1}^n$ is generated independently from the  distribution of $\x_{t-1} \mid \y_{1: t-1}$ with mean $\x_{t-1}^{a}$ and covariance $\bfSigma_{t-1}^a$ at time $t-1$. Then, if  $ \mathcal{M}_t$ is approximated by $\mathcal{\hat M}_t$, the forecast and analysis states of the EnKF satisfy 
			\begin{align}	&p^{-1}\mathbb{E}\{\|\hat\x_{t}^{f}-\x_{t}^{f}\|^2|\y_{1:t-1} \}=n^{-1}p^{-1}\operatorname{tr}(\bfSigma_{t}^f+\Xi_{\xi_t})+p^{-1}\|\mu_{\xi_t}\|^2,\label{eq:enkf1stepmsefm}\\
				&p^{-1}\mathbb{E}\{\|\hat\x_{t}^{a}-\x_t^a\|^2|\y_{1:t}\} =p^{-1}\operatorname{tr}\big(n^{-1}(\I-\K_t\H_t)\bfSigma_{t}^f\big)+V_t
    \label{eq:enkftstepmseam}
    \\&=O\Big(n^{-1}+p^{-1}\|(\I-\K_t\H_t)\mu_{\xi_t}\|^2+
    p^{-1}\|\y_t-{\H}_t \x_{t}^f\|^2\nn \big(\|\Xi_{\xi_t}\|^2+\tilde\Delta_{\K,t}\big)\Big)
 	\end{align}
  where $\tilde\Delta_{\K,t}=\E\|\tilde\K_t-\K_t\|^2$ and $\tilde\K_t$ is the estimated $\K_t$  with true model $\mathcal{M}_t$.
	  \end{theorem}

The theorem suggests that the model approximation error creates extra  terms related to $\mu_{\xi_t}$ and
$\|\bfXi_{\xi_t}\|$, which are absent from the results in Theorem \ref{enkfconsit_1stephigh}  where accurate model is assumed. Specifically,  the assimilation MSEs for fixed $p,q$ and  $p,q\rightarrow \infty$ are, respectively,    
\begin{align} 
     \label{eq:hd1stepMmu}
\mathbb{E}\{\|\hat\x_{t}^{a}-\x_t^a\|^2|\y_{1:t}\} &\asymp \max\{n^{-1},
\; \|(\I-\K_t\H_t)\mu_{\xi_t}\|^2,\;\|\Xi_{\xi_t}\|^2\} \\
p^{-1}\mathbb{E}\{\|\hat\x_{t}^{a}-\x_t^a\|^2|\y_{1:t}\} &\asymp \max\{n^{-1},
\; p^{-1}\|(\I-\K_t\H_t)\mu_{\xi_t}\|^2,\;p^{-1}{q}\tilde\Delta_{\K,t},\;p^{-1}{q}\|\Xi_{\xi_t}\|^2\}.\nn
\end{align} 
If $\|\Xi_{\xi_t}\| \neq 0$ but $\mu_{\xi_t}=\0$, which means $\mathcal{\hat M}_t$ is  unbiased to $ \mathcal{M}_t$,  then 
for fixed  $p$ and $q$,
\begin{equation}
    \label{eq:fix1stepM}
    \mathbb{E}\{\|\hat\x_{t}^{a}-\x_t^a\|^2|\y_{1:t}\}=O\big(n^{-1}+\|\y_t-{\H}_t \x_{t}^f\|^2(\|\Xi_{\xi_t}\|^2+n^{-1})\big)=O\big(n^{-1}+\|\Xi_{\xi_t}\|^2)\big).
\end{equation} 
Thus, if $\|\Xi_{\xi_t}\|^2=O(n^{-1})$, 
 then $\mathbb{E}\{\|\hat\x_{t}^{a}-\x_t^a\|^2|\y_{1:t}\}=O(n^{-1})$, that is, the model approximation error does not alter the  order of magnitude of  
 the assimilation MSEs of the EnKF. 
However, if 
$n\|\Xi_{\xi_t}\|^2 \rightarrow \infty$, then 
$\mathbb{E}\{\|\hat\x_{t}^{a}-\x_t^a\|^2|\y_{1:t}\}=O(\|\Xi_{\xi_t}\|^2)$, which means the model approximation error dominates the assimilation MSE. 
When  $p,q\rightarrow \infty$,  $ p^{-1}\mathbb{E}\{\|\hat\x_{t}^{a}-\x_t^a\|^2|\y_{1:t}\} \asymp \max\{n^{-1},
\; p^{-1}{q}\tilde\Delta_{\K,t},\;p^{-1}{q}\|\Xi_{\xi_t}\|^2\}.$

  If $\mathcal{\hat M}_t$ is a biased approximation to $\mathcal{M}_t$ so that $\mu_{\xi_t}\neq 0$, { the approximation bias would also contribute to the MSE. Specifically,  }
for either fixed $p$ and $q$ or  $p,q\rightarrow \infty$,  the  MSEs 
in the  analysis states of the EnKF are, respectively,
\begin{align} 
     \label{eq:hd1stepMmu}
\mathbb{E}\{\|\hat\x_{t}^{a}-\x_t^a\|^2|\y_{1:t}\} &\asymp \max\{n^{-1},
\; \|(\I-\K_t\H_t)\mu_{\xi_t}\|^2,\;\|\Xi_{\xi_t}\|^2\} \text{ and }\\
p^{-1}\mathbb{E}\{\|\hat\x_{t}^{a}-\x_t^a\|^2|\y_{1:t}\} &\asymp \max\{n^{-1},
\; p^{-1}\|(\I-\K_t\H_t)\mu_{\xi_t}\|^2,\;p^{-1}{q}\tilde\Delta_{\K,t},\;p^{-1}{q}\|\Xi_{\xi_t}\|^2\}.\nn
\end{align}

The above analysis shows that for both the fixed and diverging dimensions and both the unbiased and biased approximation of $\mathcal{M}_t$, the approximation errors 
 as represented by   $\mu_{\xi_t}$ and  $\|\Xi_{\xi_t}\|^2$ contributes substantially to the assimilation MSE, which suggests that a good approximation of the model operator is a significant aspect of data assimilation. 
While this may be obvious, Theorem \ref{enkfconsitM_1step} describes exactly how it happens.

The accumulation of the model approximation error is more pronounced in the $h-$step assimilation MSEs as shown in Theorem \ref{enkfconsitM_tstep} and  more details of the $h-$step assimilation in the presence of the model error can be found in Theorem S.3.4 of the SM.  
Define 
  $ D_{t,h}=\max_{t< k \leq t+h}{\left\|\I-\check \K_{k} \H_{k}\right\|}$, where  ${\check\K}_{k}=(\bfSigma_{k}^f+\Xi_{k}) \mathbf{H}_{k}^{\mathrm{T}}\big(\mathbf{H}_{k} (\bfSigma_{k}^f+\Xi_{\xi_k})\mathbf{H}_{k}^{\mathrm{T}}+\mathbf{R}_{k}\big)^{-1}$, which is the underlying  Kalman Gain matrix under the imperfect model. Denote a non-stochastic constant  $C_{t,h}^D=(1-L_{t,h}^{h}D_{t,h}^{h})/(1-L_{t,h}D_{t,h})$.
   \begin{theorem}	\label{enkfconsitM_tstep}
		Suppose Assumptions \ref{ass1}(i) or \ref{ass1}(ii),  \ref{ass2}, 
   \ref{ass3}, \ref{ass4} hold  and   the analysis ensemble $\{\x_{t,j}^a \}_{j=1}^n$ is generated independently from the  distribution of $\x_{t} \mid \y_{1: t}$ with mean $\x_{t}^{a}$ and covariance $\bfSigma_{t}^a$ at  time $t$. If  $ \mathcal{M}_t$ is approximated by $\mathcal{\hat M}_t$, by repeatedly performing the ensemble forecast-analysis process  $h$ times, at  time $t+h$, the forecast and analysis states of the EnKF satisfy
     \begin{align}
	& p^{-1}\mathbb{E}\{\|\hat\x_{t+h}^{f}-\x_{t+h}^{f}\|^2|\y_{1:t+h-1} \}
   \nn\\  &
				= O\Big( n^{-1}+p^{-1}\|\mu_{\xi_{t+h}}\|^2
			+p^{-1}L_{t,h}^2(C_{t,h-1}^{D})^2\max_{t< k\leq t+h-1}\big\{\Delta_{\check\K_{k}}\| \y_{k}-{\H}_{k} \x_{k}^f-\H_{k}\mu_{\xi_{k}}\|^2\big\}\Big),  \nn \\
				&p^{-1}\mathbb{E}\{\|\hat\x_{t+h}^{a}-\x_{t+h}^{a}\|^2|\y_{1:t+h}\}	
   	\nn\\&= O\Big( n^{-1}+
  p^{-1}\|(\I-\K_{t+h}\H_{t+h})\mu_{\xi_{t+h}}\|^2+p^{-1}\|\Xi_{\xi_{t+h}}\|^2\|\y_{t+h}-\H_{t+h}\x_{t+h}^f\|^2\nn
			\\&\quad+p^{-1}(C_{t,h}^{D})^2\max_{t< k\leq t+h}\big\{\Delta_{\check\K_{k}}\| \y_{k}-{\H}_{k} \x_{k}^f-\H_{k}\mu_{\xi_{k}}\|^2\big\}\Big). \nn 
    \end{align}
  \end{theorem}
 
  Compared with the result in Theorem \ref{enkfconsit_tstephigh} under the correct model, there are extra terms related to $D_{t,h}=O(\|\I-\K_{k}\H_{k}\|+\|\Xi_{\xi_k}\|\|\H_{k} \|)$, $\|\mu_{\xi_{t+h}}\|$ and $\|\Xi_{\xi_{t+h}}\|$ resulting from the accumulation of the approximation error of the model operator $\mathcal{M}_t$ for the $h-$step assimilation MSE in Theorem \ref{enkfconsitM_tstep}. Compared with the one-step assimilation MSE in Theorem \ref{enkfconsitM_1step} under the imperfect models, the error term related to the approximation error of  $\mathcal{M}_t$ is inflated by $(C_{t,h}^{D})^2$. 
%
 The $h-$step assimilation MSE for the imperfect models under the fixed and high dimensions are, respectively, 
\begin{align}
&\mathbb{E}\{\|\hat\x_{t+h}^{a}-\x_{t+h}^{a}\|^2|\y_{1:t+h} \}\asymp \max\big\{\big( 1
			+(C_{t,h}^D)^2\big)n^{-1},\; \|\mu_{\xi_{t+h}}\|^2,\;\|\Xi_{\xi_{t+h}}\|^2\big\},\nn \\
&p^{-1}\mathbb{E}\{\|\hat\x_{t+h}^{a}-\x_{t+h}^{a}\|^2|\y_{1:t+h} \}\asymp \max\big\{ 
n^{-1},\;p^{-1}\|\mu_{\xi_{t+h}}\|^2,\; p^{-1}{q}\|\Xi_{\xi_{t+h}}\|^2,\;(C_{t,h}^D)^2p^{-1}{q}\Delta_{\check\K_{k}}\big\}. \label{eq:hdhstepM}
 \end{align}

Specifically, for the \hd~case, the MSEs of the EnKF analysis state $\hat\x_{t+h}^{a}$  to the Kalman Filter $\x_{t+h}^{a}$ for the $h$-step assimilation using the banding or mid-banding, tapering and thresholding estimates of $\bfSigma_t^f$ can be obtained by replacing  $\Delta_{\check\K_{k}}$ in equation \eqref{eq:hdhstepM} with the specific convergence rates 
$({n}^{-1}{\log p})^{\alpha /(\alpha+1)}$, $n^{-2 \alpha /(2 \alpha+1)}+{n}^{-1}{\log p}$ and $c_0^2(p)(n^{-1}{\log p})^{(1-\gamma)}$, respectively,
according to  \eqref{eq:kt-banding}. 
For the $h$-step assimilation, the approximation error of the model operator $\mathcal{M}_t$ has  a stronger  influence on  the MSE of the data assimilation of the EnKF resulting from the accumulation of the approximation error. 
This highlights that a good approximation of the model operator is significant to data assimilation.

In summary, the approximation error associated with the model operator $\mathcal{M}_t$ significantly affects the MSE in the assimilation process of the EnKF, particularly in scenarios involving multiple-step ahead data assimilation.
Thus, an iterative HD-EnKF algorithm is introduced in S.1.2 of the SM to account for the model approximation error to some extent since most geophysical models are imperfect in practice, of which the main idea is to substitute $\hat\x_t^{f}$ in the original forecast sample covariance matrix by the analysis state $\hat\x_t^{a}$ in the iteration.

\section{Simulation Studies}\label{simulation} 
This section reports results of simulation studies performed on two popular geophysical models, the Lorenz-96 and the Shallow Water Equation models,  to demonstrate the proposed HD-EnKF algorithms. 
These two models are often used for testing new approaches in atmospheric sciences for data assimilation  and the spatial-temporal  forecasts.  
 The proposed HD-EnKFs are compared with the standard EnKF and the inflation method  \citep{liangMaximumLikelihoodEstimation2012a} under both the correct and imperfect models.  Details of the proposed HD-EnKF algorithms are given in Section S.1 of the SM, including the consistent HD-EnKF update in S.1.1 and the iterative HD-EnKF update algorithms in S.1.2,   which are based on the consistent estimators of the forecast error covariance matrix $\bfSigma_t^f$ by the banding, mid-banding,  tapering and thresholding estimators. The iterative HD-EnKF algorithm in S.1.2 of the SM is to account for the model approximation error in the imperfect models.
\subsection{Lorenz-96 model}
Lorenz-96 model (L96; \citealp{Lorenz96})  is a continuous in time but discrete in space model for modeling spatially extended chaotic systems in the atmosphere. It is a 
nonlinear model for continuous evolution of atmospheric variables 
 such as temperature or vorticity  at   
 $p$ locations on  a latitude circle 
    defined by the 
   ordinary differential equation, for $j=1,\ldots, p$, 
\begin{equation}\label{l96}
   {d\x_t^{(j)}}/{dt}=(\x_t^{(j+1)}-\x_t^{(j-2)})\x_t^{(j-1)}-\x_t^{(j)}+F \overset{\triangle}{=}f_j(\x_t,t),\,
\end{equation}
 with $\x_t^{(-1)}=\x_t^{(p-1)}, \x_t^{(0)}= \x_t^{(p)}$ and $\x_t^{(p+1)}=\x_t^{(1)}$, where $\x_t^{(j)}$ denotes the $j^{th}$ dimension of the  state variable $\x_t$,  
$(\x_t^{(j+1)}-\x_t^{(j-2)})\x_t^{(j-1)}$ is  the nonlinear term,  $-\x_t^{(j)}$ is the damping term and $F$ is a forcing term.

L96 can be solved {approximately} via the fourth-order Runge-Kutta time integration scheme (RK4; \citealp{butcherNumericalMethodsOrdinary2016}), 
which approximates the model as $\mathcal{M}_{t+1}\left(\x_t\right)=\x_t+ h(k_1+2 k_2+2 k_3+k_4)/{6},$
where $k_1=\f\left(\x_t, t\right)$,
    $k_2=\f\left(\x_t+h k_1 / 2, t+h / 2\right)$, 
     $k_3=\f\left(\x_t+h k_2 / 2, t+h / 2\right)$, 
     $k_4=\f\left(\x_t+h k_3, t+h\right)$
and $\f = \left(f_1,f_2,\dots,f_p\right)^{\T}$. 
An approximation to the true state was simulated by (i) implementing the RK4 method with a finer time step, say $h=0.05$;
and (ii) adding a random noise at each time step so that $   \x_{t+1} = \mathcal{M}_{t+1}\left(\x_t\right) + \w_{t+1}$.  
In our experiment, $\w_t\sim \mathrm{N}(0,\sigma_0\I_p)$ and $\sigma_0$ were set as $0.1,0.25$ and $0.5$, respectively. 
We experimented $p=40$ and $100$, under which the system behaves chaotically, since the L96 model behaves like chaos for $p \geq 7$ \citep{bedrossianRegularityMethodLower2022}. 
The true forcing was $F=8$. 
We also considered the misspecified model setting by making 
$F=6,7,9,10$ in the assimilation processes, 
corresponding to the imperfect models in Section \ref{sec-5-imperfect-model}. 
The initial state was
$x_1^{(j)}=F$ for $j\neq 20$ and $x_1^{(20)}=F+0.001$, and  the initial forecast ensemble was generated by $\x_{1,j}^f=\x_1+\N_p(0,0.1\I_p)$ for $j=1,\ldots,n$.

The 
 observations $\y_t$ were made 
at $q$ randomly selected  components of the state vector $\x_t$  
 with $q=30,40$ and $100$ while $p=40$ and $100$, respectively, 
followed by adding observation errors $\bfep_{t}\stackrel{IID}\sim \N_q(0,\R_t)$ to the true states. 
The $(i,j)-$element of $\R_t$ was $R_{ij} = 0.5^{\min\left\{|i-j|, q-|i-j|\right\}}$, which prescribes spatially correlated observation errors. 
The observation operator $\H_t$ was a $q\times p$ matrix with the $(j,i)$ element being $1$ if the $j-$th component of $\y_t$ was the observed $i-$th dimension of 
the state $\x_t$. 
We simulated observations every $4$ 
time steps for a total of $2000$ steps 
and reported the results for the last $1000$ steps to allow the system to settle down. 
The ensemble size $n$ was $20,30,60$ and $90$. 
All the experiments were repeated 500 times.

We considered the HD-EnKF schemes using the banding, mid-banding, tapering and thresholding covariance estimators, which were compared with the standard EnKF that utilizes the sample covariance 
and the inflation methods \citep{liangMaximumLikelihoodEstimation2012a}. 
The performances of these methods were  evaluated by the root-mean-square error (RMSE) of the analysis state $\hat\x_t^a$ 
to (i) the underlying analysis state $\x_t^a$ by the Kalman Filter utilizing the true covariances and (ii)  the true state $\x_t$. 
To gain accurate estimations of $\bfSigma_t^f$ for the oracle 
analysis $\x_t^a$, we used the standard EnKF 
with a very large ensemble size $1000$ (relative to $p$ in the simulation), and  used   the sample covariance
to replace $\bfSigma_t^f$ in the Kalman Filter update formula to obtain 
$\x_t^a$. 

\begin{figure}[!ht]
    \centering
    \includegraphics[width = 1\textwidth]{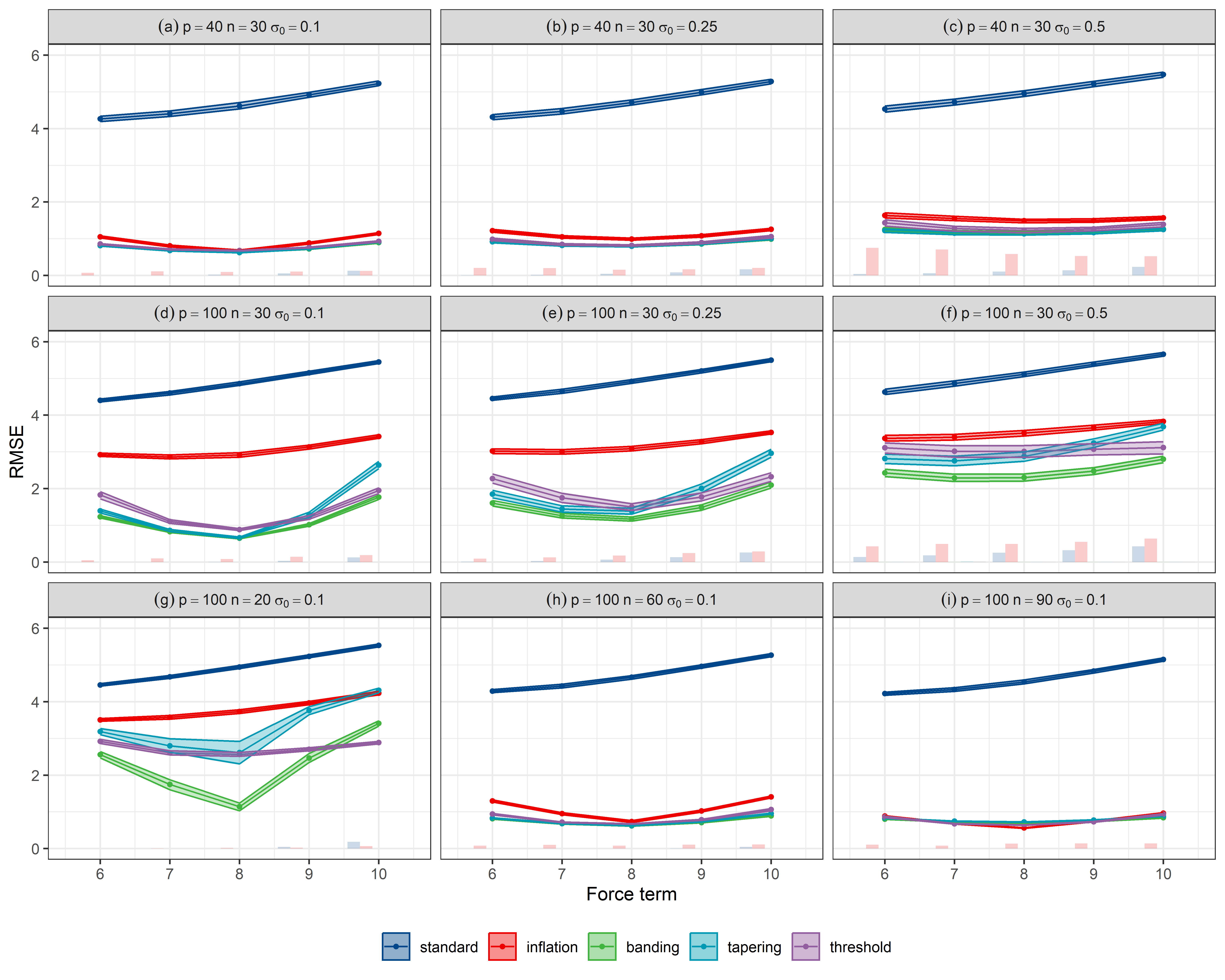}
    \caption{Average RMSEs to the oracle analysis state and their $25-75\%$ quantile bands (color bands)  as a function of forcing $F$ used in ${\cal{M}}_t$ (true value is 8) by the standard EnKF (blue), the modified EnKF  methods: the inflation (red), banding (green), tapering (cyan) and thresholding (purple) for $p=q=40$ and $100$, $n = 20,30,60,90$, $\sigma_0 = 0.1, 0.25$ and $0.5$, respectively. 
    The light-colored bars in the lower part of each panel denote the divergence rates of the assimilation schemes. 
    }
    \label{fig:L96-RMSE-alln}
\end{figure}

Figure \ref{fig:L96-RMSE-alln} displays  the average RMSEs to the oracle analysis state of the five assimilation methods, namely the  standard EnKF,
the inflation method and the three HD-EnKFs equipped with the banding and tapering (both respect the circular bandable structures),  and the thresholding estimators, under different forcing $F$  at moderate dimension $p=40$ and high dimension $p=100$ with different  $\sigma_0$ and $n$. 
It shows that 
the HD-EnKF with the three high dimensional 
 covariance estimators  
had much smaller RMSEs than the standard EnKF and the inflation methods 
at both the correct ($F=8$) and misspecified ($F\ne 8$) models. 
Even when $F$  deviated more from the true value $8$,  the HD-EnKFs still achieved analysis states much closer to the oracle than the inflation method, although both strains of methods showed increased RMSEs with respect to $|F - 8|$.
The standard EnKF with the sample covariance estimator had much larger assimilation errors in all cases. 

Comparing Figure \ref{fig:L96-RMSE-alln}'s panels (d)-(f) with panels  (a)-(c),  the advantages of the proposed HD-EnKFs over the inflation method were more apparent for the higher dimension $p=100$.  The analysis RMSEs of the HD-EnKFs kept at a 
lower level especially when the force $F$ was equal or close to the true value ($F=8$), while the analysis RMSE of the inflation method ranked behind but better than the standard EnKF. This means that the HD-EnKFs are quite attractive as the geophysical models in earth science are of ultra high dimension.
The figure also displays the percentages of divergence of the five methods. The inflation method endured quite high divergences which increased substantially with the model error (e.g. reached  50\% when $\sigma=0.5$), even higher than the standard EnKF. In contrast, the three HD-EnKF methods had quite low divergence rates for all cases. 

 Panels (d), (g)-(f) of Figure \ref{fig:L96-RMSE-alln} demonstrate the influence of the ensemble size on the analysis RMSEs. Although the analysis RMSEs of  the inflation and the HD-EnKF methods all became 
 smaller with the increase of the ensemble size $n$, the proposed HD-EnKFs outperformed the inflation method, 
 especially when the ensemble size $n$ was small (panels (d) and (g) with $n=20$).
 This suggests that  the proposed  
 methods could provide more accurate analysis states when the ensemble size was not big enough. Ensemble size is a significant restriction for the large geophysical models as they are extremely costly to run. 

\begin{table}[ht]
\centering
 \caption{Average  RMSEs to the oracle analysis state and true state of the five EnKF schemes and their divergence rates (in parentheses)
    for Lorenz-96 model 
    with $p=40$, $q=30$, $n=30$, $\sigma_0=0.1$, and $F=8$ for the correct and other $F$ values for misspecified models.  
    }
       \resizebox{0.8\textwidth}{!}{
\begin{tabular}{llllllll}
  \hline
 & method & F=6 & F=7 & F=8 & F=9 & F=10 \\ 
  \hline
 & standard & 4.26 (0.01) & 4.43 (0.06) & 4.69 (0.15) & 5.05 (0.29) & 5.41 (0.54) \\ 
 & inflation & 1.32 (0.80) & 0.96 (0.65) & 0.77 (0.63) & 1.03 (0.69) & 1.38 (0.86) \\ 
oracle & banding & 1.10 (0.11) & 0.90 (0.08) & 0.83 (0.09) & 0.99 (0.12) & 1.27 (0.29) \\ 
 & tapering & 1.08 (0.13) & 0.87 (0.07) & 0.79 (0.06) & 0.99 (0.13) & 1.31 (0.25) \\ 
 & threshold & 1.12 (0.06) & 0.90 (0.02) & 0.84 (0.04) & 1.00 (0.04) & 1.35 (0.16) \\ 
 \hline
 & standard & 4.28 (0.01) & 4.44 (0.06) & 4.70 (0.15) & 5.05 (0.29) & 5.40 (0.54) \\ 
 & inflation & 1.42 (0.80) & 1.04 (0.65) & 0.84 (0.63) & 1.11 (0.69) & 1.47 (0.86) \\ 
 true & banding & 1.23 (0.11) & 1.01 (0.08) & 0.93 (0.09) & 1.09 (0.12) & 1.37 (0.29) \\ 
& tapering & 1.21 (0.13) & 0.98 (0.07) & 0.90 (0.06) & 1.09 (0.13) & 1.42 (0.25) \\ 
& threshold & 1.24 (0.06) & 1.00 (0.02) & 0.93 (0.04) & 1.09 (0.04) & 1.44 (0.16) \\ 
   \hline
\end{tabular}
}\label{tab:L963}
\end{table}

Table \ref{tab:L963} displays the analysis RMSEs of different methods and their divergence rates under sparser observations 
 $q=30$ while $p=40$, which illustrates the influence of the sparsity in the observations on the assimilation accuracy. Although the analysis RMSEs of different assimilation methods were slightly higher than those under more observations $p=q=40$, the RMSEs of the HD-EnKF methods were smaller than those of the other methods for imperfect models ($F\neq 8$) and were comparable to the inflation method for the correct model ($F= 8$). However, it is noted that the inflation method had a higher divergence rate which reached 63\% even under the correct model ($F= 8$) while the proposed methods had low divergence rates, showing their robustness to the sparsity of the observations.

\subsection{Shallow Water Equation model}
The second experiment was conducted on the
Shallow Water Equation model (SWE). 
The SWE is a system of hyperbolic/parabolic partial differential equations governing fluid flow, which is 
  widely used in river flow simulations, as well as in other settings where the water depth is significantly lower than the horizontal length scale of motion. 
The barotropic nonlinear SWE takes the following form \citep{Lei2009AHE}, for $0 \leq z_1 \leq L, 0 \leq z_2 \leq D,$
\begin{equation}\label{swe}
    \begin{gathered}
        \frac{\partial u}{\partial t}+u \frac{\partial u}{\partial z_1}+v \frac{\partial u}{\partial z_2}-f v=-g \frac{\partial h}{\partial z_1}+k \nabla^2 u, \\
        \frac{\partial v}{\partial t}+u \frac{\partial v}{\partial z_1}+v \frac{\partial v}{\partial z_2}+f u=-g \frac{\partial h}{\partial z_2}+k \nabla^2 v, \\
        \frac{\partial h}{\partial t}+u \frac{\partial h}{\partial z_1}+v \frac{\partial h}{\partial z_2}=-h\left(\frac{\partial u}{\partial z_1}+\frac{\partial v}{\partial z_2}\right)+k \nabla^2 h,
    \end{gathered}
\end{equation}
where  the fluid velocity components $u$ in $z_1$ direction and $v$ in  $z_2$ direction as well as the fluid depth $h$ are the model variables, 
$g$ is the gravity acceleration $9.8 \mathrm{~m} \mathrm{~s}^{-2}$, 
$f$ is the Coriolis parameter defined as constant $10^{-4} \mathrm{~s}^{-1}$, 
$k$ is the diffusion coefficient
specified as  $5\times 10^4 \mathrm{~m}^2 \mathrm{~s}^{-1}$, and 
$L$ and $D$ are the domain limits set as $500 \mathrm{~km}$ and $300 \mathrm{~km}$, 
 respectively.  
The SWE was discretized with a uniform grid spacing of $10 \mathrm{~km}$ in the $z_1$ and $z_2$ directions and solved 
by the Lax-Wendroff method with a time step of $30$ seconds. %
Thus, the SWE model 
had a dimension of $p^{\prime}=50 \times 31$ for each state variable $u,v,h$, leading to 
the state vector $\x_t=(u_t,v_t,h_t)^{\T}$ of dimension $p=3p^{\prime}=4650$.
The initial height (depth) was given by $      h(z_1, z_2)=  H_0+H_1  \tanh \big({9 (D / 2-z_2)}/2/D\big)+H_2 \operatorname{sech}^2\big(9 (D / 2-z_2)/{D}\big)  \sin \big({2 \pi}z_1/{L} \big),$
where $H_0, H_1$ and $H_2$  were set to be $50\mathrm{~m}, 5.5 \mathrm{~m}$ and $3.325 \mathrm{~m}$, respectively. 
The initial velocity field was derived from the initial height field with the geostrophic relation. 

The SWE was simulated for $48$ hours with a total of 5760 time steps at 30 second intervals 
after a half hour (60-steps) pre-running.  
The 
observations of $u, v$ and $h$ were made at $q^{\prime}=310$ 
grid points from $10$ randomly selected rows in each direction every 3 hours, making
the whole observation vector $\y_t=(u_t,v_t,h_t)^{\T}$ be of dimension $q=3q^{\prime}=930$, which means the observations were quite sparse with $q=0.2p$. 
To generate the imperfect models, the diffusion coefficient $k$ was also considered as $10^4 \mathrm{~m}^2 \mathrm{~s}^{-1}$ in the assimilation process.
The observation error variances were 
$\sigma_{\mathrm{u}}^2=\sigma_{\mathrm{v}}^2=0.5, \sigma_{\mathrm{h}}^2=1.0$
for the three dimensions and $\R_t=diag(\sigma_{\mathrm{u}}^2\I_{p^{\prime}},\sigma_{\mathrm{v}}^2\I_{p^{\prime}},\sigma_{\mathrm{h}}^2\I_{p^{\prime}})$. 
The ensemble size $n$ was 100, and the initial ensemble
was generated by adding random noises from $\N_p(0,\I_p)$ to the initial state $\x_1$. 

\begin{figure}[!ht]
    \centering
    \includegraphics[width = 1\textwidth]{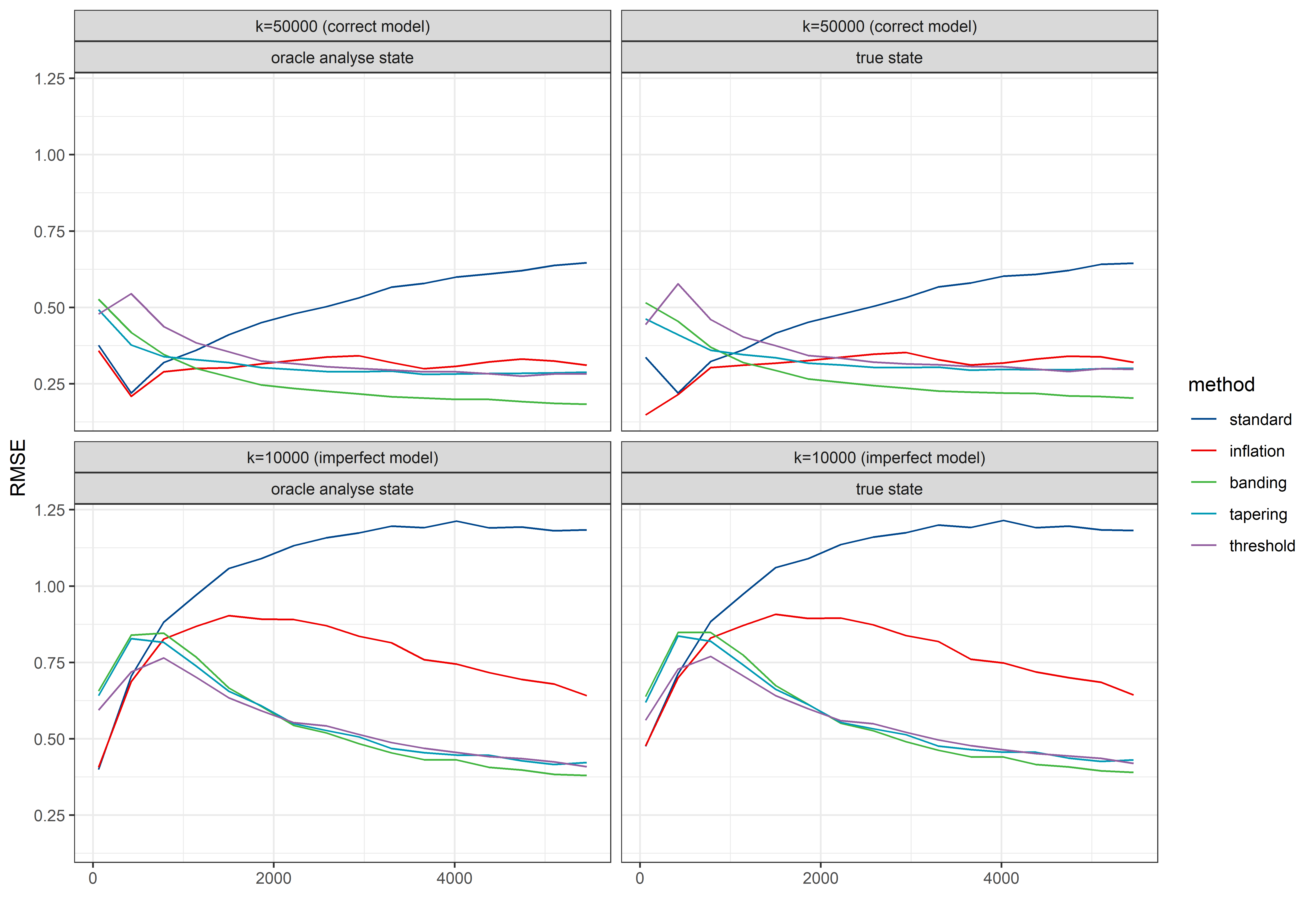}
    \caption{RMSEs to the oracle analysis state (left panels) and the true state (right panels) at each assimilation step by the standard, inflation, banding, tapering and threshold under the correct model ($k=5\times 10^4$) and imperfect model ($k=10^4$) with $n=100$, $p=4650$ and $q=930$ on the SWE model.}
    \label{fig:SWE-RMSE2}
\end{figure}

\begin{table}[ht]
    \centering
    \caption{Average RMSEs to the oracle analysis state and true state over the last 24 hours (2880 steps) and their $25\%$ and $75\%$ quantiles (in parentheses) by the five EnKF schemes for the SWE model under the correct  ($k=5\times 10^4$) and the imperfect  ($k=10^4$) models with $n=100$, and $p=4650$ and $q=930$. 
    }
     \resizebox{0.8\textwidth}{!}{
    \begin{tabular}{lcccc}
        \hline
        \multirow{2}*{method} & \multicolumn{2}{c}{k = 50000 (correct model)} & \multicolumn{2}{c}{k = 10000 (imperfect model)} \\ 
        \cline{2-5}
        ~ & oracle & true & oracle & true \\ 
        \hline
standard & 0.61 (0.45, 0.76) & 0.62 (0.47, 0.78) & 1.16 (1.05, 1.26) & 1.18 (1.07, 1.28) \\ 
  inflation & 0.34 (0.27, 0.40) & 0.38 (0.31, 0.44) & 0.75 (0.68, 0.82) & 0.77 (0.70, 0.84) \\ 
  banding & 0.20 (0.17, 0.22) & 0.26 (0.23, 0.28) & 0.42 (0.40, 0.45) & 0.46 (0.43, 0.48) \\ 
  tapering & 0.25 (0.23, 0.28) & 0.30 (0.28, 0.33) & 0.44 (0.41, 0.46) & 0.47 (0.44, 0.50) \\ 
  threshold & 0.28 (0.26, 0.30) & 0.33 (0.31, 0.35) & 0.44 (0.42, 0.46) & 0.48 (0.46, 0.50) \\ 
        \hline
    \end{tabular}
    }
    \label{tab:swe}
\end{table}
Figure \ref{fig:SWE-RMSE2} reports the RMSEs at each assimilation time step of {the five EnKF methods (the standard EnKF, inflation and three HD-EnKF)}
to the oracle analysis states $\x_t^a$ 
and the true states $\x_t$  
under the correct  ($k=5\times 10^4$) 
and the imperfect ($k=10^4$) models while Table \ref{tab:swe} summarizes the average analysis RMSEs 
 over the last 24 hours (2880 steps), together with their  $25\%$ and $75\%$ quantiles. 
Figure \ref{fig:SWE-RMSE2}  
shows the three proposed HD-EnKF methods outperformed the inflation and the standard EnKF methods for both the correct and imperfect models with much smaller RMSEs after the system settled down and that the superiority became more apparent over time, especially under the misspecified model.  Specifically,  the RMSEs at each assimilation time step of the three proposed HD-EnKF schemes became lower and lower over time compared to increased RMSEs for the standard EnKF for all cases and  for the inflation method under true models. 

It is obvious from Table \ref{tab:swe} that the HD-EnKF methods had significantly lower RMSEs and narrower 
25\%-75\% quantile bands with more evident superiority under imperfect models. Specifically, the proposed HD-EnKFs could reduce the RMSE by 41\% compared to the inflation method, implying that the proposed methods could attain more accurate and robust assimilated results even under extremely sparse observations and high dimensional states. 
Thus, the proposed methods are highly appealing, especially considering the high dimensional geophysical models in earth science.
\section{Discussion}\label{sec-discussion} 

In this paper, we propose HD-EnKF algorithms to suit the needs of high dimensional data assimilation. We take the opportunity to study the theoretical
properties of the EnKF under both the fixed and the high dimensional
cases for both the correct and misspecified models. 
 The study shows that the proposed HD-EnKF can replace the widely used inflation method for data assimilation of a large dynamic system based on two considerations. One is that the established theoretical analysis on the consistency and the error bounds makes the proposed HD-EnKF methods more understandable with a theoretical guarantee; and the other is the better algorithm convergence and smaller RMSEs than the inflation method as shown in the simulation studies. The latter means that a smaller ensemble size for the EnKF can be used at a given level of accuracy which is particularly attractive as the running of the geophysical models is generally very expensive.  
 The proposed HD-EnKF has great potential for applications in earth science where the geophysical models, the state variables or the observations are high dimensional, such as global carbon calculation, ocean data assimilation, weather forecasting, sea ice prediction and the generation of high-resolution data sets.
 	
	

\bigskip
\begin{center}
{\large\bf SUPPLEMENTARY MATERIAL}
\end{center}

We provide detailed steps of the proposed HD-EnKF algorithm, bandwidth selection of the mid-banding estimator,  additional theoretical results and simulation results, and the details for the proofs of the
theoretical results in the Supplementary.

\bibliographystyle{chicago}
\bibliography{refe}       





\end{document}